# Multiple Asteroid Systems: Dimensions and Thermal Properties from Spitzer Space Telescope and Ground-Based Observations*

F. Marchis[a,g], J.E. Enriquez[a], J. P. Emery[b], M. Mueller[c], M. Baek[a], J. Pollock[d], M. Assafin[e], R. Vieira Martins[f], J. Berthier[g], F. Vachier[g], D. P. Cruikshank[h], L. Lim[i], D. Reichart[j], K. Ivarsen[j], J. Haislip[j], A. LaCluyze[j]

a.  Carl Sagan Center, SETI Institute, 189 Bernardo Ave., Mountain View, CA 94043, USA.
b.  Earth and Planetary Sciences, University of Tennessee 306 Earth and Planetary Sciences Building Knoxville, TN 37996-1410
c.  SRON, Netherlands Institute for Space Research, Low Energy Astrophysics, Postbus 800, 9700 AV Groningen, Netherlands
d.  Appalachian State University, Department of Physics and Astronomy, 231 CAP Building, Boone, NC 28608, USA
e.  Observatorio do Valongo/UFRJ, Ladeira Pedro Antonio 43, Rio de Janeiro, Brazil
f.  Observatório Nacional/MCT, R. General José Cristino 77, CEP 20921-400 Rio de Janeiro - RJ, Brazil.
g. Institut de mécanique céleste et de calcul des éphémérides, Observatoire de Paris, Avenue Denfert-Rochereau, 75014 Paris, France
h.  NASA Ames Research Center, Mail Stop 245-6, Moffett Field, CA 94035-1000, USA
i.  NASA/Goddard Space Flight Center, Greenbelt, MD 20771, United States
j.  Physics and Astronomy Department, University of North Carolina, Chapel Hill, NC 27514, U.S.A

* Based in part on observations collected at the European Southern Observatory, Chile Programs Numbers 70.C-0543 and ID 72.C-0753

Corresponding author:
Franck Marchis
Carl Sagan Center
SETI Institute
189 Bernardo Ave.
Mountain View CA 94043 USA
fmarchis@seti.org

**Abstract:**
We collected mid-IR spectra from 5.2 to 38 μm using the Spitzer Space Telescope Infrared Spectrograph of 28 asteroids representative of all established types of binary groups. Photometric lightcurves were also obtained for 14 of them during the Spitzer observations to provide the context of the observations and reliable estimates of their absolute magnitudes. The extracted mid-IR spectra were analyzed using a modified standard thermal model (STM) and a thermophysical model (TPM) that takes into account the shape and geometry of the large primary at the time of the Spitzer observation. We derived a reliable estimate of the size, albedo, and beaming factor for each of these asteroids, representing three main taxonomic groups: C, S, and X. For large (volume-equivalent system diameter $D_{eq}$ >130 km) binary asteroids, the TPM analysis indicates a low thermal inertia ($\Gamma \leq$ ~100 J s$^{-1/2}$K$^{-1}$m$^{-2}$) and their emissivity spectra display strong mineral features, implying that they are covered with a thick layer of thermally insulating regolith. The smaller (surface-equivalent system diameter $D_{eff}$ <17 km) asteroids also show some emission lines of minerals, but they are significantly weaker, consistent with regoliths with coarser grains, than those of the large binary asteroids. The average bulk densities of these multiple asteroids vary from 0.7-1.7 g/cm$^3$ (P-, C- type) to ~2 g/cm$^3$ (S-type). The highest density is estimated for the M-type (22) Kalliope (3.2 ± 0.9 g/cm$^3$). The spectral energy distributions (SED) and emissivity spectra, made available as a supplement document, could help to constrain the surface compositions of these asteroids.

## 1. Introduction

The population of known multiple asteroid systems has grown steadily over the past decade thanks to the development of photometric lightcurve survey campaigns, high angular resolution imaging from the ground using adaptive optics (AO), the Hubble Space Telescope (HST), and radar observations. The ~190 known multiple asteroid systems exhibit a wide diversity in key characteristics, including the separation of the components, their size ratios, their mutual orbits, and the locations of the systems. This diversity strongly implies that they did not all form in the same manner, and that they most likely evolved differently.

Using the current sample of known binary asteroid systems, several investigators (Pravec & Harris, 2007; Descamps & Marchis, 2008) introduced simple nomenclatures for multiple asteroid systems based on known physical parameters, such as:

- The size of the primary component of the system (or the equivalent size of both components for a similar-size binary), which can be approximated from its visible magnitude (assuming a geometric albedo $p_v$ of ~0.07 for C-group asteroids and ~0.25 for S-group asteroids), or estimated directly from mid-IR observations. If the asteroid has a close passage with Earth or is sufficiently large, radar observations can also directly determine the size of the components.
- The relative size of the components, which can be determined by the modeling of mutual events detected in their lightcurves, by direct stellar occultation timing, and/or the brightness ratio (assuming identical albedos).
- The distance of the secondary which can be directly measured on high-angular resolution observations collected with AO and HST or radar observations.
- The orbital period of the components of the systems, which can be derived from lightcurve observations.

**Large asteroids with small satellites**, or Type-1 (T1) in Descamps & Marchis (2008) or group L in Pravec and Harris (2007). These are multiple asteroid systems with a large primary $D_p$>90 km and one or several small satellites ($D_s$<$D_p$/5) in circular or low eccentricity orbits ($e$<0.3) with $a/R_p$<15 ($a$, the semi-major axis of the mutual orbit; $R_p$, $D_p$ the radius and diameter of the primary, $D_s$ the diameter of the satellite). The satellites could be the secondary products of an instability after successive spin-ups of the pulverized primary (Descamps and Marchis, 2008). They could also have been formed from ejecta from a large impact (Durda et al. 2004). The satellite(s) of the thirteen well-characterized T1/L multiple systems can be imaged directly AO on ground-based telescopes and with the Hubble Space telescope.

**Similar size double asteroids, or Type-2** (T2) in Descamps & Marchis (2008) or group B in Pravec and Harris (2007). These are similar size $(D_s \sim D_p)$, synchronous, double asteroids with $a/R_{eq} \sim$ 3-8 ($R_{eq}$: radius of a sphere with the same volume as the multiple system), having a circular orbit around their center of mass. The large ($D_s \sim D_p \sim 85$ km) binary T2 asteroid (90) Antiope can be imaged directly with AO systems, but the other ten ($D_p \sim 10$ km) well-characterized T2/B have been discovered by lightcurve observations, which have a large amplitude and display the typical, so called "U-V shape", signatures of mutual events (eclipse/occultation) between the components of the system (see Behrend et al. 2006). Descamps & Marchis (2008) showed that they spread along the same equilibrium sequence of two fluids approximated by two identical ellipsoids rotating around their center of gravity (the "Roche" ellipsoids).

**Small asynchronous systems,** part of the Type-3 (T3) of Descamps & Marchis (2008) or group A of Pravec and Harris (2007). These are small binary systems with $D<10$ km and a mass ratio <0.2, found in the near-Earth asteroid (NEA) and inner main-belt populations by lightcurve photometric surveys (Pravec et al. 2006) or radar observations (Ostro et al. 2006). As described in Jacobson & Scheeres (2011a), this binary asteroid population is part of an evolutionary track driven by "rubble pile" geophysics. The secondary might be synchronized (as in the case of 1999KW4, Ostro et al. 2006), the mutual orbit could be eccentric, or the system could be composed of two moons (triple/ternary asteroid systems). This sub-population of multiple systems is one of the most difficult to classify.

**Contact-binary asteroids** are a group (type-4 or T4) defined in Descamps & Marchis (2008) but are part of the B class in Pravec & Harris (2007). They are detected through their high-amplitude lightcurves (e.g., (3169) Ostro, [Descamps et al. 2007]), radar observations (Benner et al., 2006) and some of them can be directly imaged with AO (e.g., (624) Hektor primary; Marchis et al. 2006c). Descamps & Marchis (2008) mentioned that these dumbbell-shape asteroids could be the members of the same evolutionary scenario, mimicking the rotational fission of a rapidly spinning fluid mass. In the Jacobson & Scheeres (2011a) evolutionary track they are the result of the reaccretion of components of similar sizes from a double system affected by the binary YORP (BYORP) process.

**Small wide binaries** or loosely-bound binary system are introduced as the W-group in Pravec & Harris (2007). They are small binary systems ($D \sim 4$-8 km) detected by direct imaging with AO or HST and which have a companion orbiting at $a/R_{eq} > 20$. The mutual orbits of most of them are unknown and the formation scenario is unclear. They may be ejecta from large asteroidal impacts that remained gravitational linked due to their similar velocity (see Durda et al. 2004).

The present work describes the analysis of mid-infrared spectra collected with the Infrared Spectrograph (IRS) on the Spitzer Space Telescope of several binary asteroids belonging to these sub-groups. Thermal-IR observations are a powerful means of determining asteroid sizes, albedos and surface properties. Our work was motivated by the need to determine the sizes and albedos of these asteroids to derive their average densities (if the mass is known), the sizes of the components, and their taxonomic classes. In Section 2, we describe how we collected and processed observations with Spitzer/IRS to extract the spectra of 28 multiple asteroid systems, as well as a photometric lightcurve survey conducted in parallel. In section 3 we use a thermal model to derive the sizes, albedos, beaming factors and their associated errors for these asteroids. Section 4 presents thermophysical modeling of seven asteroids for which we have a good estimate of their shapes. In Section 5 we describe the spectral emissivity features that are associated with the composition and grain size of the surface, but also the size of the asteroids. In Section 6, we review the sizes, density, and porosity of each system based on our data. The conclusion summarizing the main findings of our work is found in Section 7.

## 2. Observations and Data Processing

### 2.1 Targets and Observations

In 2007, the Spitzer telescope and its InfraRed Spectrograph (Houck et al. 2004) was the best facility to record high quality thermal-IR spectra of asteroids. To this purpose, we compiled a list of known binary systems from Noll (2006) and Richardson and Walsh (2006), and added those that were recently discovered by photometric surveys. We rejected objects that were not reasonably observable with IRS in Spitzer observing cycle 4 for reasons of solar elongation or brightness, as well as all objects that had been observed with IRS already. This resulted in a total of 26 selected targets. Their diameters (estimated from their absolute optical magnitudes H) vary from 1 to 250 km. For the brightest ones (45, 107, 121, 283), we selected the IRS high-resolution mode with R~600 in two modules (SH; 9.9-19.5 μm & LH; 18.8-37.1 μm) to avoid saturation. For the other ones we used the low-resolution mode (R~61-127) resulting in spectra consisting of two short-wavelength long-slit segments (SL1; 7.4-14.2 μm, SL2; 5.2-8.5 μm,) and two long wavelength long-slit segments (LL1; 19.5-38.0 μm, LL2; 14.0-21.5 μm). For our faintest targets, we added a timing constraint to ensure that they were observed when bright enough to give a good signal to noise ratio (SNR). The IRS pixel scale is 1.8" or coarser, so none of these binary systems can be spatially resolved.

We obtained 19.91 h with the Spitzer/IRS to observe these 26 binary asteroids (PI: F. Marchis, program ID 40164) from 5.2 to 38 μm, and the program was executed in cycle 4 from August 28, 2007 to May 28, 2008. Three spectroscopic observations of asteroids: (3671) Dionysus, (3982) Kastel, (35107) 1991VH failed because i) the target was fainter than anticipated, or ii) the target was passing nearby a bright star at the time of Spitzer observation.

We completed our sample by searching and retrieving Spitzer/IRS observations of binary asteroids available in the Spitzer Data Archive[1]. We found useable observations of (87) Sylvia, (130) Elektra, and (762) Pulcova recorded in Cycle 0 (PI: D. Cruikshank, program ID 88 & 91) and observations of (22) Kalliope and (216) Kleopatra in Cycle 2 (PI: L. Lim, program ID 20481). Table 1 contains the observing circumstances of all these Spitzer/IRS data.

The observed 28 targets, 23 main-belt asteroids and five near-Earth asteroids, and the characteristics, such as their estimated size based on the H value (assuming an albedo of 0.23 or 0.04, depending on their taxonomic class) and/or IRAS observations (SIMPS catalog by Tedesco et al. 2002) are listed in Table 2. This work presents 28 thermal-IR spectra of multiple asteroids including:

- Nine T1 including 3 triple asteroids with an estimated diameter (*D*) between 125 and 261 km
- Six T2 with *D* from 1 to 12 km
- Nine T3 with *D* estimated between 1 and 8 km

---

- Two T4 (suspected binary asteroids) with $D$ =11 and 15 km
- Two W with $D$ = 6 km

We also looked for the taxonomic classes of these asteroids extracted from several surveys such as SMASS (Bus and Binzel, 2002) and S3OS2 (Lazzaro et al. 2004) in the visible and Birlan et al. (2011) in the near-infrared. Since 2009, we have been also conducting our own spectroscopic survey in both the visible (Reiss et al. 2009) and near-infrared (Marchis et al. 2011a) to help determine the spectral types of various asteroids. Using those classifications, our Spitzer/IRS target list contains three asteroids from the C-complex, 8 from the X-complex, and 14 S-complex objects, including three V-type asteroids. Three asteroids have still an unknown taxonomic classification.

**[Insert Tables 1 & 2 here]**

## 2.2 Spitzer/IRS data reductions

The data reduction of the Spitzer/IRS data is fully described in Emery et al. (2006). Much of the basic data processing (e.g., dark subtraction, flat field correction, correction of several IRS specific artifacts, flux calibration) is performed by the Spitzer Science Center (SSC). We therefore started our processing with the Basic Calibrated Data provided by the SSC pipeline. We then applied the same data reduction method described in Lim et al. (2011). We removed the background emission, due mostly to zodiacal clouds, by subtracting frames from the two dither positions. We corrected bad pixels using the IRSCLEAN software[2] provided by the SSC, and performed a regular extraction of 1D spectra using SPICE[3], also supplied by the SSC. For several asteroids, multiple observations were recorded per "module" (equivalent to a wavelength band range). In these cases, we took the median of all cycles to produce a final spectrum for each segment. The different wavelength segments were scaled relative to each other using the flux intensities in the wavelength regions of overlap. The entire spectrum was rescaled to the average flux of the LL2 segment in the case of observations in the low resolution mode and to the average of the scale factors for all segments in the case of observations in the high resolution mode. The scaling factors (80-95%) fell within the 20% per order photometric flux uncertainties of the Spitzer/IRS (Emery et al. 2006).

The Spitzer/IRS spectra in the 5-38 μm region for each asteroid shown in Fig. 1a-g is dominated by thermal emission. It peaks between 15 and 25 μm, depending mostly on the asteroid's heliocentric distance, with a peak intensity between 0.014 and 32 Jy depending on the size of the asteroid and its distance to the Sun. From an estimated thermal spectrum derived for the target closest to the Sun and with a high albedo, we determined that the contribution of the solar reflected flux is less than 10% at < 6 μm, and negligible at longer wavelengths.

**[Insert here Fig 1a-1g]**

---

## 2.3 Lightcurve Observations

Multiple asteroid systems can be seen in mutual event geometry (eclipse, occultation or transit), which could have an impact on the shape and intensity of their integrated thermal spectrum. These kinds of scheduled observations are of great interest if we want to measure directly the thermal inertia, as was done for the Trojan double asteroid (617) Patroclus and Menoetius (Mueller et al. 2010). The main goal of our work here was, however, to provide a direct measurement of the equivalent diameter of the binary system, hence its density if the mass can be reasonably estimated.

In coordination with our Spitzer program, we conducted an intensive visible light photometric survey to give the rotational context of the Spitzer observations. The photometric lightcurve of a binary asteroid recorded in visible light gives indications of the shapes of the components (with a typical amplitude Δmag between 0.1 and 0.3) and their spin periods (typically 3-10h). If mutual events can be observed, the lightcurve will also display attenuations with a period corresponding to the revolution of the moon around the primary and with a profile and intensity that vary with time due to the relative motion between the asteroid and the observer around the Sun. More than 100 asteroids are known to possess a satellite based on the study of their lightcurves, which show variable signatures of mutual events.

In our case, we recorded photometric observations near the times of the Spitzer/IRS observations to check if the system was observed in transit/occultation, in which case the calculated effective diameter ($D_{eff}$) will not correspond to the equivalent diameter of both components. Additionally, for some of our targets, we used these photometric measurements to derive an accurate estimate of the absolute magnitude $H_V(\varphi)$. The absolute magnitude is a function of the rotation phase of the primary $\varphi$. We define the absolute magnitude during the Spitzer observations as $H_V(\varphi=0)$. We decided to use the spin of the primary as a reference for our Spitzer observations, since it is the most likely cause of a photometric variation in the lightcurve if i) the asteroid is a T1 binary (the primary elongated shape is dominating the photometric variation, ii) the asteroid is a T2 binary observed outside mutual event configurations (the spin of the two component elongated shapes which are tidally locked ($P_{spin} = P_{orbit}$) is the predominant cause of photometric variations), iii) the asteroid is a T3 binary observed outside the mutual event configurations, in which case the spin of the primary is the dominant variable since the size ratio between the satellite and the primary ($D_s/D_p$) for stable systems varies from 0.1 to 0.5 (Jacobson & Scheeres, 2011b).

We identified in our survey several lightcurves that are useable to provide the rotational context of the Spitzer observations for 14 asteroids that are members of the T2, T3 and T4 binary classes. The characteristics of these lightcurves (telescopes, date of observations, assumed color) are listed in Table 3. Observations from telescopes of size 0.6-1.6m in three observatories, including two in the southern hemisphere (Cerro Tololo, Chile, and Pico dos Dias, Brazil) and one in the northern hemisphere (Lick Observatory, CA), were used for our photometric survey in the standard R and I photometric bands. For each target, we display in Fig. 2a-n their lightcurves phased to the period of the primary ($P_{prim}$) (extracted from the Virtual Observatory Database of Binary Asteroids, or VOBAD database[4] and listed in Table 3). We observed a range of photometric variations in these lightcurves from 0.09 to 1.15 mag with an average on the sample of 0.47 mag. These photometric variations are significantly larger than the assumed error in the absolute magnitudes of these asteroids (~0.1 *mag*), demonstrating that we can indeed improve our analysis by determining the absolute magnitude of the system at the time of the Spitzer observations. Because these lightcurves were recorded 0.5 to 87 days before/after the Spitzer observations, and one month after/before the Spitzer observations on average in our sample, we can neglect the change in geometry . The typical change in solar phase is <2 deg between the times of our Spitzer and groundbased observations, so less than the difference in phase due to the relative position of Spitzer telescope with respect to Earth (see below).

---

[4] http://cilaos.berkeley.edu/PHP_scripts/VOBAD/VOBAD_portal.html

**[Insert here Table 3]**

Table 4 contains a list of absolute magnitudes for those systems used in this work. Four different strategies were used to derive the accurate absolute magnitudes and their errors:

- For 10 targets for which we measured high SNR lightcurves, we calculated the absolute magnitude in our data for the same rotational phase as the Spitzer observations ($H_V(0)$). To this purpose we extracted the flux of the asteroid and compared it to nearby photometric stars in the field of view by aperture photometry to derive its apparent magnitude. Knowing the distance to the Sun, the distance to Earth, and the solar phase at the time of the photometric observations and using the *H-G* magnitude system described in Bowell et al. (1989), we derived the absolute magnitude $H_V(0)$. In practice, we assumed a *G* parameter of 0.15 (see Table 4) and we estimated the error in the photometric extraction by averaging the absolute magnitude over a rotational phase range up to ±0.1. If several lightcurves were used to derive the absolute magnitude, we derived a final measurement by averaging them and adding quadratically the 1-sigma variation with the error from the rotational phase. In this case we were also able to derive the G parameter. The final absolute magnitudes and errors, both listed in Table 4, encompass the measurement errors, and also a possible poor approximation of the G parameter. The difference between our measured ($H_V(0)$) and the JPL HORIZONS absolute magnitude is up to 1 *mag* for (809) Lundia, and ~0.6 *mag* for (1333) Cevenola, (1509) Esclangona, and (3782) Celle, most likely because these T2 and T3 systems exhibit large magnitude variations and/or have a poorly known absolute magnitude estimate. By using in our analysis $H_V(0)$ instead of the ephemeris H value in the thermal modeling in Section 3, we correct for possible mutual effects between both components of the system. The diameter that we derived for these asteroids is the effective diameter ($D_{eff}$) of a sphere with the same projected area as the binary system at the time of the Spitzer observation. By combining lightcurve observations of the system taken over several years and thus different geometries, to build a 3D shape model of a binary system (e.g. Descamps et al. 2007; Scheirich & Pravec, 2009) it will be possible to convert directly these measured effective diameters into equivalent-diameter ($D_{eq}$ diameter of a sphere with the same volume than the combined components) and derive the size/shape of each component. In Section 4, we present this analysis for seven T1 asteroids for which the shape of the primary is estimated by combining several years of photometric observations and the thermal flux of the satellite is negligible due to its small size ($D_s/D_p$~0.03-0.05).

- Four targets had low SNR and/or disparate lightcurves. For three of these - (5905) Johnson, (4674) Pauling, 1999 HF1 - we used the JPL HORIZONS absolute magnitude. For (3749) Balam, we used measurements of *H* and *G* from Polishook et al. (2011). To take into account the elongated shapes of the primary objects and the geometry of these binary system, we estimated the errors on the absolute magnitudes by measuring the 1-sigma variations on a large rotational phase range (>±0.5) from our lightcurves. This 1-sigma error is relatively small (<0.12 *mag*) and comparable to the 1-sigma error on the absolute magnitudes provided by the ephemeris.

- We did not collect lightcurves for two targets (1999AW1 & 2000DP107) because of their faintness or their location in the sky at the time of the Spitzer campaign. In this case we used the JPL HORIZONS absolute magnitude and assumed an error of 0.1.

- For the nine large T1 asteroids, we also used the JPL HORIZONS absolute magnitudes and assumed an error of 0.1. These asteroids are bright (V<13) and have often been observed since their discoveries at the end of the 19[th] Century, we can assume that their absolute magnitudes have been well calibrated. In Section 4, we present a more accurate data analysis using thermophysical models for seven of these asteroids, taking into account their shapes and geometries of the primary at the time of the Spitzer observations.

Because the Spitzer telescope is located in an Earth-trailing orbit, its relative position with respect to Earth varies with time (by about 0.1 AU per year). In this work, we cannot calculate the effect of the change in geometry for 10 asteroids for which we do not possess shape models, but we can provide estimates of the effect on the basis of the lightcurves. Table 1 shows that the difference in phase angle is on average 5 deg and always less than 12 deg, corresponding to 1-3% of a rotation of the primary component. After inspecting the lightcurves shown in Fig. 2a-n, we estimate that the difference in viewing geometry between Spitzer and the Earth translates to an average difference in magnitude of only 0.03 (corresponding to 0.14 km) therefore negligible with respect to the typical error on the size measurements of these ~10 km asteroids (1-2 km). As already mentioned, a careful analysis, including shape modeling of the components of the system, is needed to reduce the effect of this error.

**[INSERT Table 3 & Figure 2a-n]**

## 3. Thermal Modeling

### 3.1 Method

The usual temperatures of NEAs and main-belt asteroids cause their thermal fluxes to peak in the mid-IR. At a distance $\Delta_{AU}$ of ~0.5 AU from the observer, and 1 AU from the sun, a small asteroid of ~1 km with albedo ~0.1 will have a flux ~25 mJy. Therefore, the use of Spitzer & IRS was ideal for our studies.

The simplest model to describe the thermal emission from asteroids is the Standard Thermal Model (STM; Lebofsky et al., 1986), which assumes a non-rotating asteroid of spherical shape with a solar phase angle of zero, and a thermal inertia of zero, implying a night-side temperature of T = 0 K and thus no emission. Thus the total flux, $F_\lambda$, is calculated from the integral over the area of the sunlit hemisphere of the Planck-function, $B(\lambda, T)$ by

$$F_\lambda = \frac{1}{\Delta^2} \int \varepsilon B(\lambda, T) \cos\theta_e \, dA \tag{1}$$

where ε is the bolometric emissivity (assumed value of 0.9; Emery et al. 2006 and references therein). The flux varies with the cosine of the emission angle, $\theta_e$ since the surface is assumed to be Lambertian. For the STM case, the temperature distribution at the surface of the asteroid is modeled by

$$T(\phi) = T_{max} \cos^{1/4}\phi \quad for \quad 0 \le \phi \le \pi/2 \tag{2}$$

as described by Lebofsky and Spencer (1989), $\phi$ is the subsolar / Earth angle and $T_{max}$ is the subsolar temperature, which is in turn described by Emery et al. (2006) as,

$$T_{max} = \left[ \frac{S_0(1 - A_B)}{r_{AU}^2 \sigma \varepsilon \eta} \right]^{1/4} \tag{3}$$

where $S_0$ is the solar flux at 1 AU (1366 W/m$^2$ ; Frohlich 2009), $r_{AU}$ is the object heliocentric distance, $A_B$ is the bolometric bond albedo, $\sigma$ is the Stefan-Boltzmann constant, and $\eta$ is the beaming factor.

The bolometric bond albedo is related to the geometric albedo, $p_V$, by $A_B = p_V \times q$, where the phase integral $q = 0.290 + 0.684 \times G$ with $G$ the slope parameter listed in Table 4 (Bowell et al., 1989; Fowler and Chillemi, 1992).

The scaling factor $\eta$ was introduced by Jones and Morrison (1974) to account for the departures from the assumption of zero thermal inertia and the anisotropy of the thermal emission towards the sunward direction. This latter effect is commonly termed "beaming" and is due to surface roughness. Lebofsky and Spencer (1989) found $\eta \sim 0.756$ empirically by calibrating it to the large main-belt asteroids Ceres and Pallas.

However, this model proved to be not as accurate for NEAs, and thus Harris (1998) used a modified approach called Near-Earth Asteroid Thermal Model (NEATM), where η is used as a free parameter to improve the fit to the data. A second difference between NEATM and the STM is that whereas the STM integrated emitted flux over the entire sunlit hemisphere and applied a linear phase correction of 0.1 mag/degree, NEATM models the phase correction by integrating over the hemisphere viewed by the observer, only part of which is illuminated by sunlight. Using NEATM, several works (Delbo et al. 2003, Wolters et al. 2005, Harris et al. 2009 and references there in) have found typical η values from ~1 to 1.5 with larger values for increasing solar phase angle.

Binzel et al. (2002) showed that the rotation and shape characteristic of NEAs were similar to main-belt asteroids with $D < 12$km. Marchis et al. (2008a, 2008b) also showed that for the T1 binary asteroids the NEATM model provides an equivalent size close to the resolved measurements from AO observations. We therefore use NEATM for initial thermal modeling of the Spitzer data. The radius and geometric albedo are left as free variables, allowing us to find the best fit to the data by minimizing the $\chi^2$.

The geometric albedo is related to the effective diameter (Pravec and Harris, 2007)

$$D_{eff} = \frac{1329}{\sqrt{p_v}} 10^{-H_v/5} \qquad (4)$$

with $H_V$, the absolute magnitude of the asteroid described in the previous section (Table 4).

Due to the excellent signal-to-noise of the Spitzer data, this method works very well and produces a good fit to the data, as shown in Fig. 1. The resulting parameters, equivalent diameter ($D_{eff}$), geometric albedo ($p_V$), beaming factor ($\eta$), and the 3-sigma error (see Section 3.2) of the 28 targets of our sample are listed in Table 5.

## 3.2 Results and error estimate

Our sources of error are similar to those identified in Emery et al. (2006). To compute the uncertainties on the derived parameters for each target, we developed a Monte-Carlo method for which we generated a new spectrum slightly different from the observed one by varying its integrated flux within the 10% uncertainty and the brightness per wavelength of the spectrum within its 3-sigma uncertainty. This part of the Monte-Carlo simulation takes into account the observational errors introduced by the instrument and data processing. We also considered the error introduced by the H-G asteroid magnitude system by varying the absolute magnitudes Hv of the asteroids within the error indicated in Table 4. The random numbers were generated assuming a uniform distribution with a new absolute magnitude within the range of the ±3-sigma errors listed in Table 4. In the case of (3749) we also included a variation of the $G$ parameter considering the value and error derived by Polishook et al. (2011) ($G$=0.40 ± 0 .02) for this asteroid.

We ran 100 trials per target and performed the analysis in the same manner as for the observed spectra. Table 5 summarizes the 3-sigma error derived from our Monte-Carlo simulation. The error in the size estimate ($D_{eff}$) is 12% on average (with a maximum 15%) and only 6% for the beaming factor. The error in the albedo is significantly larger, with an average of 34%. Our Monte-Carlo simulation showed that the error in the albedo is mostly due to the uncertainty in the absolute magnitude $H_v$ of the asteroids. Refined measurements provided by future all-sky surveys like Pan-Starrs or LSST will help reduce the error in the albedo.

**[Insert here Table 5]**

## 4. Thermophysical Modeling

Where possible, we used a detailed thermophysical model (TPM; see Sect. 4.1) to analyze our Spitzer data. As this model provides a realistic description of thermal processes on asteroid surfaces, it would be expected to lead to more accurate diameter and albedo estimates than the simpler models discussed above (see also Müller et al., 2005 and Mueller et al, 2006, for TPM diameters of spacecraft targets (25143) Itokawa and (21) Lutetia, respectively, obtained pre-encounter, which agree with spacecraft results to within a few percent). Additionally, knowledge of thermal properties, chiefly thermal inertia, $\Gamma$, can be gained. In turn, meaningful thermophysical modeling requires previous knowledge of the object's shape and spin state, which is currently unavailable for most asteroids. As described in Sect. 4.2, this information is available for seven of our targets: (22) Kalliope, (45) Eugenia, (87) Sylvia, (107) Camilla, (130) Elektra, (121) Hermione, and (283) Emma.

## 4.1 Model description

Our TPM is based on previous work by Spencer (1990) and Lagerros (1996, 1998). It takes direct account of the object shape and spin state, and provides for a realistic physical description of thermal conduction into the subsoil (leading to thermal inertia) and surface roughness (leading to thermal beaming). This TPM has been used in the analysis of thermal-IR observations of a number of asteroids (e.g. Harris et al., 2005, 2007; Mueller et al., 2006, 2010). A detailed description can be found in Mueller (2007).

To summarize, the object shape is approximated as a mesh of triangular facets, typically derived from the inversion of optical lightcurves. The thermal emission is modeled as the sum of the emission of the facets that are visible to the observer, weighted by their projected area. On each facet, the temperature is calculated as a function of the insolation during one rotation period, by numerically integrating the 1D heat diffusion equation for assumed values of $\Gamma$.

Lateral heat conduction can be neglected due to the small penetration depth of the diurnal heat wave, well below the resolution of typical shape models. Surface roughness is modeled by adding craters of circular cross-section to the facets; shadowing and mutual heating of facets (through radiative exchange of energy) is fully taken into account. Since it is typically not feasible to disentangle the observable effects of thermal inertia and roughness from observations at a single epoch, we adopt four different roughness models (see Table 6) that run the gamut of plausible roughness values.
**[Insert Table 6]**

## 4.2 Shape models

Models of shape and spin state were downloaded from the DAMIT online data base (see Durech et al., 2010, for a description) on Oct. 28, 2010. Shape models on DAMIT are based on the inversion of optical lightcurves and are updated as more data become available. All shape models used in this study are convex (like most shape models derived from inversion of optical lightcurves). This technique and these models were validated by comparison with direct imaging observations with AO and occultations.

The model of (22) Kalliope is derived from Kaasalainen et al. (2002) and is consistent with the pole direction and stellar occultation analysis from Descamps et al. (2008). The model for (45) Eugenia was originally published by Kaasalainen et al. (2002) and validated by AO comparison by Marchis et al. (2008a). The model of (87) Sylvia, derived from Kaasalainen et al. (2002) is consistent with AO images (Marchis et al. 2006b). The model for (107) Camilla is an update of that presented by Torppa et al. (2003) based on additional lightcurve data by Polishook (2009) and validated by AO comparison by Marchis et al. (2008a). For (121) Hermione, the DAMIT model is based on lightcurves and direct AO observations given in Descamps et al. (2009). The model of (130) Elektra is derived from Durech et al. (2007) and agrees well with a 2010 occultation (Durech et al., 2010) and AO images (Marchis et al. 2006b). The model for (283) Emma is an update of that presented by Michalowski et al. (2006) using additional unpublished lightcurve data by M. Fauerbach. Two solutions for spin and shape are allowed by the lightcurve data of Emma, we use both. Spin-state parameters for all TPM targets are given in Table 7.

### 4.3 Other input parameters

Spitzer-centric ephemerides of our seven TPM targets were downloaded from JPL HORIZONS for the time of our observations. TPM input parameters are $r_{AU}$ and $\Delta_{AU}$ (Table 1) as well as the heliocentric and Spitzer-centric ecliptic target coordinates. Observation times were corrected for the light travel time between target and Spitzer. The subsolar and sub-Spitzer latitudes on the targets are given in Table 7, as well as the local time at the sub-Spitzer point. (87) Sylvia, (121) Hermione, (130) Elektra, and, to a lesser extent, (107) Camilla were observed at nearly equatorial aspect, but the aspect angles of (22) Kalliope, (45) Eugenia, and (283) Emma were significant.

As is customary in the thermal modeling of asteroids, objects are assumed to be gray bodies with emissivity $\varepsilon = 0.9$. Prior to fitting, IRS data were binned, typically to a total of 27 wavelengths per object, in order to keep the computational time within reasonable limits.

**[INSERT Table 7]**

### 4.4 Results and Discussion

After initial runs over wider and coarser grids, we adopted thermal inertia values between 5 and 250 J s$^{-1/2}$K$^{-1}$m$^{-2}$ with a step width of 5 J s$^{-1/2}$K$^{-1}$m$^{-2}$.

For each thermal-inertia value and roughness model (see Table 6), TPM fluxes were calculated, and best-fit $D_{eq}$ and $p_V$ values were found through $\chi^2$ minimization. Typically, for any given roughness model the $\chi^2$ ($\Gamma$) curves (e.g., Fig. 3) have clear but distinct minima at different $\Gamma$ values. This is due to the strong degeneracy between the observable effects of thermal inertia (which to first order decreases day-side temperatures) and roughness (increasing day-side temperatures), rendering the two very difficult to disentangle from single-epoch observations such as ours. However, the corresponding best-fit diameters and albedos depend very little on roughness or thermal inertia—this behavior is familiar from published TPM work. The true uncertainty in thermal inertia and roughness is dominated by systematic effects due to, e.g., simplifications made in the thermal modeling and/or uncertainties in the adopted models of shape and spin state.

While these are hard to quantify, we adopt conservative uncertainties of 10% in $D$ and 20% in $p_V$ (see, e.g., Mueller, 2007; Mueller et al., 2010). By comparison, the uncertainty in absolute magnitude $H_v$ can be neglected for these well-studied objects. Final results are given in Table 8, see Fig. 1 for plots of best-fit TPM continua superimposed on the data and NEATM fits.

**[INSERT here  Table 6,7]**

When comparing TPM diameters to NEATM diameters, it is important to note that the latter are a measure of the instantaneous area-equivalent diameter, while the former is intrinsically a "lightcurve average" diameter in the case of these T1 binary asteroids.  In the case of significantly non-spherical shape, significant differences between the two can arise.  In order to facilitate a meaningful comparison, we have generated synthetic TPM lightcurves of our targets at the time of the Spitzer observations and at a wavelength of 10 μm. A first-order lightcurve correction of the NEATM diameters is obtained through multiplication by the square root of the flux ratio between lightcurve average and that at the time of observations (projected area $\alpha\, D^2$, hence the square root); the resulting correction factors are given in Table 8. After this lightcurve correction, NEATM and TPM diameters of all seven TPM targets agree within the error bars (note, also, that the quoted uncertainties on the NEATM diameters only include the statistical uncertainty, while the quoted TPM results include an additional 10% systematic uncertainty). This work emphasizes the importance of correcting for lightcurve effects in accurate determinations of diameters and albedos as we did for the ten additional T2, T3, and T4 asteroids of our sample (see Section 2.3).

**[insert here Figure 3 and Table 8]**

The two models of (283) Emma's spin and shape fit the Spitzer data equally well, with practically indistinguishable results. We can therefore not further constrain (283) Emma's spin axis. This is somewhat unexpected: frequently, such spin ambiguities arise because optical observations cannot differentiate between morning and evening hemispheres. Temperature, however, displays a prominent asymmetry between morning and afternoon due to thermal inertia. For example, Harris et al. (2007) could reject, based on thermal observations, one possible spin orientation of 1998 WT24 in favor of another; their finding was later confirmed by Busch et al. (2008) based on radar observations. In the case of our Spitzer observations of (283) Emma, however, both spin solutions place the sub-Spitzer point on the afternoon side.

While this study allows only coarse conclusions to be drawn on the thermal inertia of our individual targets, the values found are typically $\Gamma \leq \sim 100$ J s$^{-1/2}$K$^{-1}$m$^{-2}$, thus rather similar to large main-belt asteroids (see Delbo & Tanga, 2009, for a recent overview) or to the similarly sized Trojan binary system (617) Patroclus-Menoetius (Mueller et al., 2010). This indicates the presence of a thick (at least cm scale) layer of thermally insulating material on the surface. This is expected for old surfaces, which have had ample time to develop a fine mature regolith. Our finding therefore implies that the surface age of our sample of large main-belt binaries cannot be distinguished from that of large main-belt asteroids as a whole.

This contrasts with near-Earth asteroids (with typical diameters in the km range): binary systems in that population display an elevated inertia (Delbo' et al. 2011), indicating relatively young surfaces that are depleted in regolith. As argued by Delbo' et al., their surfaces were rejuvenated during the binary formation event, probably due to rotational fission following radiative spin-up. In the light of this, our thermal-inertia result supports the idea that large main-belt binaries formed in a substantially different way than their much smaller counterparts in the near-Earth population.

## 5. Spectral features in the mid-IR

### 5.1 Extraction and characteristics of the emissivity spectra

The mid-IR region of the spectrum has a great potential for remote characterization of the surface of asteroids since the most intense features, referred to the Si-O stretching region, occur between 8.5 and 12.0 μm (Salisbury 1983). We computed an emissivity spectrum for each asteroid by dividing the measured SED by the modeled thermal continuum calculated in Section 3 (Fig. 1a-1h). Three emissivity features; i) the Christiansen peak (~9.1 μm) related to the mineralogy and grain size, ii) the reststrahlen bands (9.1 to 11.5 μm) produced by vibrational modes of molecular complexes, iii) two transparency features (11.5-18 μm and 18-28 μm due to scattering features of fine particles, have been detected on surface of asteroids in the main-belt (Barucci et al. 2008) and in Jupiter Trojan population (Emery et al. 2006) thanks to the great sensitivity of the Spitzer/IRS instrument. More recently, various attempts to characterize the surface composition of asteroids with featureless visible spectra, such as (22) Kalliope (Marchis et al. 2008c) and (21) Lutetia (Vernazza et al. 2011), suggested a composition similar to enstatite chondrites. Lim et al. (2011) emphasized the need for a such broad wavelength spectroscopic analysis to understand the chemical nature of (956) Elisa, a V-type asteroid. Its mid-IR emissivity spectrum recorded with Spitzer/IRS shows that the surface material has an olivine-diogenitic composition, which is not apparent in the visible/NIR (0.5 to 2.5 μm) spectrum.

Due to the low SNR of their SED, the emissivity spectra of (1333) Cevenola, (9069) Hovland, and (76818) 2000RG79 are not considered in this section (See Table S1). The interpretation of the 25 remaining emissivity spectra is complicated by the presence of an instrumental artifact known as the "SL14 teardrop" whose intensity depends in a unpredictable way upon both the source position on the slit and the extraction aperture and appears on the two dimensional SL1 spectra[5]. A gray box, located between 13.2 and 15 μm highlights the area of the asteroid spectra which could be affected by this artifact is labeled in Fig 1a-h. We improved the SNR on the emissivity spectra by convolving them with a Gaussian kernel delivering emissivity spectra with a spectral resolution of ~60 at 20 μm (right-hand panels of Fig 1a-h).

## 5.2 Discussion
Because our work includes the emissivity spectra of 25 asteroids, a detailed and quantitative investigation such as that in Lim et al. (2011) or Emery et al. (2006) involving comparison with lab spectra and modeling is beyond the scope of this work. The SED spectra and the emissivity spectra are, however, made available as supplement electronic material for future analysis (Table S2a-x) Since the present investigation is the first to present a large sample of asteroid mid-IR emissivity spectra, with diverse size and variety of spectral type, we instead offer a qualitative study of the profiles of the emission spectra of those asteroids, addressing in particular the possibility of interpreting these emission features in future works.

### 5.2.1 Emissivity spectra and taxonomic class
Figure 4 displays the emissivity spectra of 25 asteroids from our sample sorted by taxonomic class after being rebinned to a spectral resolution of 50 at 20 μm. The 1-sigma error derived from the error of the SED and at the maximum of the SED is also shown for each spectrum. It is significantly smaller (spectral contrast <1%, see Table S1) than the emissions detected with a spectral contrast between 5 and 15%.

*C-complex asteroids*: Three large T1 multiple asteroids ((45) Eugenia, (130) Elektra, (762) Pulcova) are part of the C-complex group defined by Bus and Binzel et al. (2002) and DeMeo et al. (2009). The calculated albedos of these three asteroids are quite low, averaging 0.05. All three spectra have a common broad shape: i) A lowest emission near 7 μm, ii) a strong Christiansen feature and the reststrahlen band between 8.75 and 11.75 μm, iii) followed by an obvious transparency minimum.

---

The emissivity spectra of (762) Pulcova and (45) Eugenia are interestingly similar to the spectra of the Jovian Trojan asteroids (624) Hektor, (911) Agamemnon, and (1172) Aneas described in Emery et al. (2006). A strong reststrahlen band from 9.1 to 11.5 μm with a spectral contrast of 6% and a moderate transparency feature from 11.5 to 18 μm are visible. In the longer wavelength range, between 18 and 28 μm, (45) Eugenia's spectrum is characterized by its flatness, whereas (762) Pulcova and (130) Elektra have a negative slope. (130) Elektra, a G-type asteroid based on the Tholen and Barucci (1989) taxonomic classification, is the only C-complex object with a maximum peak emission at ~16 μm. These spectra suggest that the surfaces of these C-type asteroids are roughly similar in composition, but slightly different in the mixture ratio and grain size. As noted by Emery et al. (2006), these emissivity spectra more closely resemble emission spectra from cometary comae such as C/1995 O1 (Hale-Bopp) (Crovisier et al. 1997) than powdered meteorites and regolith analogs, suggesting that the surfaces of these asteroids may consist of small silicate grains embedded in a transparent matrix. More detailed analysis including comparisons with laboratory works are needed to characterize the composition of these asteroids.

*S-complex asteroids:* Thirteen asteroids of our sample belong to the S-complex, characterized by Bus and Binzel. (2002) and DeMeo et al. (2009) by a moderate to steep UV slope and 1-μm and 2-μm absorption bands. Their geometric albedos, $p_V$, average ~0.24, which is larger than the C-complex objects. We separated our 13 objects into three classes (Sq, S, and V- classes), described in Table 6 of DeMeo et al. (2009).

The emissivity spectra of all 13 objects present a minimum at 7 μm like the C-type asteroids. The rest of the emissivity spectra are significantly different from those of the C-complex. The Christiansen feature, reststrahlen and transparency features have much lower contrast. Only the S-type asteroids show the presence of the transparency feature with a low contrast varying from 11.8 to 14.2 μm. At longer wavelengths, the emissivity spectrum is relatively flat with several emission bands of spectral contrast of varying from 5 to 10%. For instance, Sq- asteroids (69230) Hermes and (3749) Balam have similar subtle bands between 15 and 35 μm, implying a similar composition and grain size. Members belonging to the V-class ((3782) Celle, (854) Frostia and (809) Lundia) of our survey have emissivity spectra slightly different from the V-type asteroid (956) Elisa (Lim et al. 2011) at short wavelengths ($\lambda$<12 um). The Christiansen and reststrahlen features are almost nonexistent. Sq-type (1509) Esclangona ($D_{eff}$ = 9 km) has the lowest albedo ($p_V$ = 0.11) of all S-complex asteroids and an emissivity spectrum very similar to the C-type (762) Pulcova. Esclangona is classified as a A- or Ld- type asteroid by Lazzaro et al. (2004). Ld-type is defined in Bus and Binzel (2002) as outlying spectral class closely associated to X- and S-complexes but with a visible spectrum which do not fit the classical definition of S- or X- types. Our work suggests that Esclangona could be made of a mixture of material of C- and S- complex asteroids. This variability in the emissivity spectra of S-complex asteroids may imply a broader diversity in composition and surface properties for these most common classes of asteroids, which in return will complicate significantly the interpretation of surface composition based on this kind of data. Before being able to provide a reliable composition interpretation of the surface of S-type asteroids from mid-IR emissivity spectra, we need to focus our attention on laboratory work with variable grain size samples.

*X-complex asteroids:* Seven multiple asteroids are members of the X-complex asteroids, which are characterized by featureless spectra with linear or medium slope in their visible/NIR reflectance (DeMeo et al. 2009). It is well known that the X-complex encompasses E-, M- and P-types of the Tholen and Barucci (1989) classification. The albedo determinations allow us to confirm the duality of this complex since we have four asteroids with $p_I$~0.04 and three asteroids with $p_I$~0.16. The morphology of emissivity spectra of the low-albedo population is quite similar to those in the C-complex. The X-type asteroids (87) Sylvia, (107) Camilla, (121) Hermione and (283) Emma have bands with similar contrast and positions to the C-type (45) Eugenia. At longer wavelength, between 18 and 28 μm, their spectra are flat like (45) Eugenia. The X-type asteroids with higher albedo: (22) Kalliope, (216) Kleopatra and 1999HF1 have different emissivity spectra at shorter wavelength (<16 μm). (22) Kalliope ($D_{eq}$ = 167 km) which is classified as an M-type asteroid in Tholen and Barucci (1989), has a spectrum with characteristics similar to C-types, but with lower contrast. The spectra of these three asteroids also have i) lowest emission near 7 μm, ii) a strong Christiansen feature and the reststrahlen band between 8.75 and 11.75 μm, iii) followed by a discernable transparency minimum. The emissivity spectrum of (216) Kleopatra ($D_{eq}$ = 152 km) is very similar to the one of (22) Kalliope at short wavelength (<16 μm). At longer wavelength the transparency features have lower contrast. 1999HF1 ($D_{eff}$ = 4.4 km) is by comparison featureless at shorter wavelengths. (22) Kalliope and (216) Kleopatra likely have a similar composition as suggested by their emissivity spectra and their similar albedo. The difference in the emissivity spectra between 1999HF1 and the large X type asteroids may be related to the grain size of surface more than a difference in composition. Additionally, Descamps et al. (2008) showed by studying the orbital characteristics of Kalliope's satellite that this asteroid could be very old (~4 *Gyrs*) so its primary's surface could have been eroded continuously, generating a thick regolith. At longer wavelength (>16 μm) these three X-type asteroids have flat emissivity spectra.

*Unknown classification:* Two asteroids with high SNR emissivity spectra do not yet have an assigned visible/NIR taxonomic class due to the absence of reported observations. Based on their measured albedos and the overall morphologies of their emissivity spectra, we can attempt to derive the nature of their surfaces.

Asteroid (4492) Debussy has $p_V$ = 0.04, which puts it in the C or X complex. Its spectrum does not mimic well any of the C-type asteroids, even though the Christiansen and reststrahlen bands have a similar contrasts, shapes and positions to those of (45) Eugenia.

2000DP107, with moderate albedo of 0.11, displays an emissivity spectrum quite different from any of the spectra in our sample, with the exception of 1999HF1 classified as a X-type asteroid. The nature of these asteroids remains unknown and requires further analysis.

**[Include Figure 4]**

### 5.2.2 Emissivity spectra and size

Laboratory measurements have shown that mid-infrared bands are detected in emission when the grain size of the sample is of the order of the observation wavelength, typically less than 45 μm in the case of the 10-μm emission reststrahlen bands (Hunt & Logan, 1972). Figure 5 displays the emissivity profiles (R=50) of our asteroid sample with respect to their equivalent diameters derived from NEATM in Section 3.

For the asteroids with diameters larger than 130 km, the reststrahlen bands between 8.75 μm and 11.75 μm and the Christiansen peak are the dominant features in the emissivity spectra. Most of the large asteroids have similar spectroscopic profiles, with subtle differences at wavelengths <15 μm, with the exception of (22) Kalliope and (216) Kleopatra, which display weak emission bands. They are the only asteroids of our sample that are classified as a M-type, so this characteristic could indeed be a signature of large M-type asteroids with composition dominated by iron-nickel. For the other asteroids, the presence of emission bands at <15 μm implies that their surfaces are made of a layer of regolith with small grain sizes (<45 μm). This hypothesis is in agreement with the low thermal inertia derived from TPM (see Table 8) for seven asteroids with shape models. Because of the high contrast of these emission lines, it should be possible to conduct a spectral analysis from a comparison with lab mixture spectra and appropriate modeling, and thereby to identify the composition and properties of the surfaces of these eight large asteroids ((87) Sylvia, (107) Camilla, (45) Eugenia, (121) Hermione, (130) Elektra, (22) Kalliope, (283) Emma, (762) Pulcova). Such an analysis was conducted by Emery et al. (2006) using mixtures of olivine and pyroxene, or enstatite and iron for Trojan asteroids. More recently, Vernazza et al. (2011) demonstrated that by modifying the sample preparation, suspending the meteorite and/or mineral grains (diameter <30 μm) in infrared-transparent KBr powder, they could reproduce the spectral behavior of these asteroids.

For smaller asteroids with diameters less than 17 km, corresponding to sixteen objects in our sample, the profiles of their emissivity spectra are significantly more diverse. Between 9.75 and 11.75 μm, emission lines can be seen for (1089) Tama, (1313) Berna, (1509) Esclangona, (4492) Debussy and (5905) Johnson, but almost no features are detectable for the others. The lack of emissivity bands in the spectra for these small asteroids at short wavelengths is most likely related to the grain size of the material on their surface that could be linked to i) their weak gravity that does not allow them to retain a thin layer of regolith, and ii) the relatively young age of those binary systems, which has not allowed them to develop an eroded layer of regolith. Binary systems among small asteroids ($D$<17 km) are thought to have formed by fission or a mass shedding process from parent bodies that are spinning at a critical rate (Scheeres 2007, Walsh et al. 2008, Jacobson and Scheeres, 2011a). The two components formed from this disruptive process should have the same composition as the parent body, but may not share the same thermophysical properties. The thick regolith layers observed on the Moon and on asteroids (Coradini et al. 2011) are formed by micro-meteorite bombardment, which is a slow process with a time scale of $10^8$-$10^9$ years (Vernazza et al. 2009). We can therefore expect these small binary asteroid systems to have significantly less dusty material on their surfaces, and therefore not have spectral emission features due to small grain size particles.

To validate this hypothesis of a relationship between size and strength of the emission bands of asteroids, additional statistical studies of emissivity spectra for non-binary asteroid of small size and medium-sized asteroids ($15<D<120$km) are necessary. Unfortunately, our sample of binary asteroids does not contain any targets with intermediate sizes. Because Spitzer/IRS is not available anymore, this kind of work will have to be conducted with a future space telescope equipped with a spectrograph, or the SOFIA telescope if its sensitivity and stability are suitable for such an investigation.

## 6. Sizes, density and porosity

The detection of a satellite around an asteroid gives the opportunity to derive the mass of the system when the semi-major axis and period of the mutual orbit have been determined. With that information combined with the equivalent diameter derived from mid-infrared observations we are able to derive the average density of the systems. It has been shown that the densities of binary asteroids vary significantly from 0.8 g/cm$^3$ for the P-type Trojan asteroid, (617) Patroclus (Marchis et al. 2006a) to 3.6 g/cm$^3$ for the dense M-type asteroid (216) Kleopatra (Descamps et al. 2011). This diversity in density and the relationship with the asteroid taxonomic class was also inferred from the handful number of asteroids visited by space missions (Barucci et al. 2011). Our new Spitzer measurements give us the opportunity to derive the bulk densities of seventeen binary systems. Table 5 gives the average densities derived from the NEATM equivalent diameters, which corresponds to the real density of a given asteroid if the components of the system are spherical. If a 3D-shape model is available, we are able to determine a more realistic density (see Table 8) for seven multiple systems since we are correcting from the elongated shape of the primary, as discussed in Section 4.4.

For eleven asteroids of our sample, the mass remains unknown, so we can only estimate the equivalent diameter of each component of the system using the size ratio derived from the lightcurve signature of an occultation/transit or from high-resolution direct imaging by assuming the same albedo for the satellite and the primary.

## 6.1 C-complex asteroids

**(45) Eugenia:** The mutual orbits of this C-type (Bus and Binzel, 2002) triple asteroid (T1/triple) system composed of a large primary and two small 7-km and 5-km size moons were derived in Marchis et al. (2010). Combining the mass of the system ($M = 5.66 \times 10^{18}$ kg) with the TPM equivalent diameter ($D_{eq} = 198 \pm 20$ km) from Table 8 we find an average density of $1.4 \pm 0.4$ g/cm$^3$, in the upper range of the density ($1.1 \pm 0.3$ g/cm$^3$) derived by Marchis et al. (2008a) using the IRAS measurements ($D_{eq} = 215 \pm 4$ km Tedesco et al. 2002).

**(130) Elektra:** The mutual orbit of this T1 binary asteroid was derived from AO observations collected and analyzed by Marchis et al. (2008b). The total mass of the system, estimated at $M = 6.57 \times 10^{18}$ kg, combined with the radiometric size from derived from TPM analysis $D_{eq} = 197 \pm 20$ km leads to a bulk density of $1.6 \pm 0.5$ g/cm$^3$, which is higher than other C-complex asteroids. (130) Elektra is classified as a G-type asteroid in Tholen and Barucci (1989) and Ch-type in Bus and Binzel (2002). (93) Minerva, a recently discovered T1/Triple that also belongs to the same taxonomic class, has been shown also to have a high density of $1.7 \pm 0.3$ g/cm$^3$, suggesting that G-type asteroids could be the densest asteroids in the C-complex (Marchis et al. 2011c).

**(762) Pulcova:** This large T1 and Cb-type (Lazzaro et al. 2004) asteroid does not have yet a shape model derived from lightcurve inversion. Its total mass, estimated in Marchis et al. (2008a), is $M = 1.40 \times 10^{18}$ kg and was estimated from the orbital parameter of the 12-26 km diameter satellite. The NEATM model gives an effective diameter $D_{eff} = 149 \pm 19$ km, very close to the IRAS measurement ($137 \pm 3$ km) from Tedesco et al (2002). From the lightcurve we find that the asteroid is moderately elongated, with a/b~1.30. Using this equivalent diameter we found a low density $\rho = 0.8 \pm 0.1$ g/cm$^3$, at the lower limit of C-type binary asteroids.

## 6.2 X-complex asteroids

We have eight asteroids belonging to the X complex defined by Bus and Binzel (2002) and DeMeo et al. (2009) in our sample. The compositions and meteorite linkage of X-type asteroids are not completely understood yet. Clark et al. (2004) studied 42 X-complex asteroids encompassing E-, M-, P- and X- from Tholen and Barucci (1989) and Xe-, Xc-, Xk, and X-class asteroids from the Bus and Binzel (2002) taxonomic system. The geometric albedo at 0.55 μm permits the classification of an asteroid into E-, M-, P- types, with E-types (possibly linked to enstatite chondrite meteorites) at the high albedo ($p_V >$ 30%) end of the range and P-types at the low end ($p_V <$ 10%). The M-types objects have a moderate albedo 10% < $p_V$ < 30% and are possibly linked to metallic meteorites. The meteorite analogs of P-type asteroids are still unknown.

**(22) Kalliope:** This binary asteroid discovered in 2001 is certainly one of the most accurately characterized asteroid multiple system. Descamps et al. (2008) determined the mutual orbit, shape of the primary and size estimate of the components by combining AO, lightcurves with mutual event (eclipse/transit/occultation) signature. We use the 3D-shape model developed by our colleagues to perform an independent TPM analysis. We derive an equivalent diameter $D_{eq}$ = 167 ± 17 km, in agreement with the size estimated by Descamps et al. (187 ± 6 km), and a moderate albedo of $p_V$ = 0.17 ± 0.03, confirming its M-type classification from Tholen & Barucci (1989). New AO observations obtained in 2010, providing a time baseline of nine years of data, combined with a new dynamical model, allowed Vachier et al. (2012) to refine the orbital elements of Linus, the companion of (22) Kalliope, and derive a total mass $7.7 \times 10^{18}$ kg, in agreement with the mass derived by Marchis et al. (2008a). By combining this new mass estimate with our independent radiometric size, we infer a density for (22) Kalliope, $\rho$ = 3.2 ± 0.9 g/cm$^3$, significantly higher than the bulk density of C-type T1 asteroids, but similar to the density of (216) Kleopatra, also an M-type asteroid (Descamps et al. 2011). The smaller size of (22) Kalliope (8% smaller than from the IRAS measurement), which was reported by Descamps et al. (2008) from a different set of observations, is a clear validation of our TPM model and the accuracy of this technique. From an early analysis of the visible/NIR reflectance spectrum and emissivity spectra, Marchis et al. (2008c), concluded that this asteroid could be made of a mixture of enstatite, olivine (forsterite) and metallic iron, with mixing ratios and grain sizes of 35% and 1.0 μm, 25% and 2.5 μm, and 40% and 1.5 μm, respectively, and with an average bulk density of ~5.0 g/cm$^3$. Considering the average density of 3.2 g/cm$^3$, the macro-porosity of Kalliope should be ~35%. The asteroid (21) Lutetia, known to have the same average density of 3.4 ± 0.3 g/cm$^3$ (Patzold et al. 2011) has a mid-IR spectrum showing emissivity signatures similar to enstatite chondrites (Vernazza et al. 2011).

**(87) Sylvia:** This T1/Triple asteroid is the first asteroid known to possess two moons, discovered in 2005 (Marchis et al. 2005a). This Cybele asteroid is also the largest member of a collisional family identified recently by Vokrouhlicky et al. (2010), implying that its two moons could be evidence of the collisional history. From the TPM model, we show that the (87) Sylvia primary is large, with $D_{eq}$ = 300 ± 30 km and a low albedo $p_V$ = 0.033 ± 0.007, suggesting that (87) Sylvia is a P-type asteroid. Using the mass measurement from Marchis et al. (2005a) ($M$= 1.48 x $10^{19}$ kg) we derive a density of $1.0 \pm 0.3$ g/cm$^3$ for this system. A similarly low bulk density for the P-type Trojan asteroid (617) Patroclus reported by Marchis et al. (2006a) ($0.8 \pm 0.2$ g/cm$^3$), and independently confirmed by Mueller et al. (2010) ($1.08 \pm 0.33$ g/cm$^3$), may indicate that the P-type bulk composition contains a significant portion of water ice or significant porosity.

**(107) Camilla:** The mutual orbit of this X-type asteroid (Bus and Binzel, 2002; Lazzaro et al. 2004) presented in Marchis et al. (2008a) gave a total mass for the system M = $1.12 \times 10^{19}$ kg. From the TPM analysis, we derive an equivalent diameter $D_{eq} = 245 \pm 25$ km and an albedo of $0.043 \pm 0.009$, suggesting that (107) Camilla belongs to the P-type class. From the mass measurement of Marchis et al. (2008a), the average density of the system is estimated as $1.5 \pm 0.3$ g/cm$^3$. This bulk density is relatively high compared to other known P-type multiple asteroid systems.

**(121) Hermione:** A refined analysis of the mutual orbit, shape and size of this T1 binary asteroid system was recently published by Descamps et al. (2009). That work also included an analysis of the same Spitzer/IRS data that we use, giving an equivalent diameter ($D_{eq} = 192 \pm 9$ km), which is close to our TPM measurement ($D_{eq} = 220 \pm 22$ km), considering the relatively large error bar. The shape of the primary is complex since AO observations showed that it is quite elongated and bifurcated (Marchis et al. 2005b). The use of TPM and the non-convex shape model derived from lightcurve and AO observations is the most appropriate means to derive the equivalent diameter of this large binary asteroid. We used a mass of $5.03 \times 10^{18}$ kg derived by neglecting the precession of the nodes due to the elongated shape of the primary, and derived a low density of $0.9 \pm 0.3$ g/cm$^3$. If we assume that the Hermione primary has a homogeneous interior, the mass of the system becomes $4.7 \times 10^{18}$ kg, leading to similar density of $0.9 \pm 0.3$ g/cm$^3$. This is comparable to the bulk density of Trojan asteroids (Marchis et al. 2006a, Mueller et al. 2010). This error bar does not take into account that the real volume may actually be 20% less than the model due to concavities, in which case the density could be slightly higher at $\rho = 0.9$ $^{+0.4}_{-0.3}$ g/cm$^3$.

**(216) Kleopatra:** This T1/triple asteroid was studied extensively in Descamps et al. (2011). These authors included an analysis of the same Spitzer/IRS data and derived an equivalent diameter ($D_{eq} = 156.4 \pm 5.8$ km), similar to our measurement within the error bar ($D_{eq} = 152.5 \pm 21.3$ km). The shape of the primary is complex since radar (Ostro et al. 2000) and AO observations (Descamps et al. 2011) showed that it is bilobated. Using a mass of $4.64 \times 10^{18}$ kg and our size measurement, we derive a high density of $2.7 \pm 1.1$ g/cm$^3$, typical of M-type asteroids. The use of a TPM model and non-convex shape will certainly improve the accuracy on the density measurement for such elongated primary asteroid. However, Descamps et al. (2011) noticed a discrepancy between the radar and the AO observations. A new shape model, combining lightcurve, AO, and interferometry, is under development (Kaasalainen et al. 2011) and could be used in the future to improve the density estimate.

**(283) Emma:** This asteroid is classified as an X-type by Tholen and Barucci (1989) and C-type by Lazzaro et al. (2004). From its mutual orbit, Marchis et al. (2008a) derived a mass of $1.38 \times 10^{18}$ kg, which translates to $\rho = 1.0 \pm 0.3$ g/cm$^3$ using the TPM with both 3D-shape models for the Emma primary. Considering the low albedo ($p_V = 0.03 \pm 0.01$), (283) Emma is most likely a P- or C-type asteroid, as suggested in Marchis et al. (2008a).

**(9069) Hovland:** The binary nature of the Hungaria asteroid (9069) Hovland was reported recently by Warner et al. (2011) from multi-component lightcurves. Their photometric study showed the primary spin period of 4.2 h, an amplitude of 0.1 mag, and an estimated orbital period of 30.2 h through the detection of mutual events. Since the geometry of the system at the time of the lightcurve observation is unknown and the secondary spin lightcurve was not detected, it is impossible to estimate the relative sizes of the components from these photometric data. Our lightcurve, recorded five days before the Spitzer observation, has a low amplitude of 0.22 mag and does not display an attenuation at the time of Spitzer observation. This suggests that the two components were separated, or that the secondary is too small to have a photometric effect on the primary lightcurve. Listed in Table 5, we derive an effective diameter $D_{eff}$= 2.9 ± 0.4 km and a geometric albedo $p_V$= 0.37 ± 0.09. In view of its high albedo, we conclude that this Xe-type asteroid (Reiss et al. 2009) in the Binzel (2002) taxonomic classification is the only known E-type asteroid (from Tholen and Barucci, 1989) measured in our sample of multiple asteroid systems.

**(137170) 1999HF1:** Pravec et al. (2002) reported 1999HF1 to be a binary near-Earth asteroid from photometric measurements that revealed two components of low amplitude (0.1-0.2 mag) and different periods (2.32 h and 14.02 h) corresponding to the spin period of the primary and the period of their mutual orbit and/or the spin of the secondary. On the assumption that the long-period component is the spin lightcurve of the secondary, or partial occultation/eclipse events, Pravec et al. (2002) derived a lower limit of 0.17 for $D_s/D_p$. The lightcurves recorded ~86 and ~87 days before the Spitzer observation shown in Fig. 2a suggest that the Spitzer/IRS observation was made when the two components of the system were not in a mutual event configuration. Therefore we can conclude that the effective diameter $D_{eff}$= 4.4 ± 0.6 km derived from NEATM model corresponds to the effective diameter of both components together. Assuming that the components are spherical, we found $D_p$ = 4.3 km and $D_s$ = 0.7 with a 3-sigma uncertainty of 0.6 km. The geometric albedo derived from NEATM ($p_V$ = 0.15 ± 0.06) is typical of M-type asteroids (comparable to the albedo of (22) Kalliope for instance). From BVRI colors, Pravec et al. (2002) concluded that the asteroid belongs to the E/M/P class from Tholen and Barucci (1989). Observations collected in the NIR with the NASA Infrared Telescope Facility (IRTF) in April 2011, confirmed that 1999HF1 could be an Xe- or Xc- asteroid in the DeMeo et al. (2009) classification. The orbit of the two components is not constrained, so the mass of the system is unknown and its density cannot be derived.

### 6.3 V-type asteroids

V-type asteroids have received significant attention over the past 10 years, because they are thought to be genetically related to (4) Vesta, one of the largest asteroids in the main belt and the first target of the *Dawn* mission. (4) Vesta and the other V-type asteroids have basaltic surface compositions with NIR spectra characterized by strong absorption bands at 1 and 2 μm. Our sample of Spitzer/IRS observations contains three binary asteroids classified unambiguously as V-type (Table 2) from their visible and NIR reflectance spectra. Since they are located in the inner main-belt ($a_{Lundia,Frostia,Celle}$<2.42 AU), in the region near (4) Vesta, they could be ejected fragments (Duffard et al. 2004) of the collision that formed the southern impact basin of (4) Vesta first observed with the Hubble Space Telescope (Thomas et al. 1997), though Celle is the only one of the three that is in the Vesta dynamical family.

**(809) Lundia:** The binary nature of (809) Lundia was suggested after an intensive photometric program conducted by Kryszczynska et al. (2009) in 2005-2007. Their lightcurves display a U-V shape typical of double binary asteroid systems in eclipse/occultation. The lightcurve recorded 20 days before our Spitzer/IRS observation, shown in Fig. 2e, has a low amplitude of 0.19 mag, implying that the system was not seen edge-on at the time of the observation.

From the NEATM effective diameter (Table 5) $D_{eff}$ = 9.6 ± 1.1 km and using the Descamps (2010) model of an inhomogeneous Roche ellipsoid interior, we derive the equivalent spherical diameters: $D_s$ = 6.4 ± 1.3 km and $D_p$ = 7.2 ± 1.4 km. These values are very close to the diameter adopted by Kryszczynska et al. (2009) even though our measured absolute magnitude is $Hv$ = 12.8 instead of $H$ = 11.8 reported in JPL HORIZONS. The calculated albedo $p_V$ = 0.14 ± 0.03 is significantly lower than the average albedo of V-type asteroids ($p_I$~0.4 for Vesta, see Tedesco et al. 2002), but comparable to the radiometrically determined albedos of two small V-types ((3908) Nyx and (4055) Magellan) measured by Cruikshank et al. (1991). Using the same theoretical study described in Descamps et al. (2010), we derived a separation between the components of $d$ = 14.6 km. With a period $P$ = 15.418 ± 0.001 $h$ of the mutual orbit, the mass is estimated as $M$ = 8.1x10$^{14}$ kg. The average density of the system of 1.77 g/cm$^3$ is very close to the density assuming a Roche ellipsoid shape. Considering the range of bulk density of HED meteorites, the adopted meteorite analogs of V-type asteroids, 2.86-3.26 g/cm$^3$ (Britt & Consolmagno, 2004; McCausland & Flemming, 2006) we infer a macroporosity for this asteroid between 40 and 50%.

**(854) Frostia**: Lightcurve observations collected by Behrends et al. (2006) revealed that (854) Frostia is a slow rotator (P~1.5713 day) with typical U-V shape characteristics of similar size binary asteroids (also called T2 binary) in mutual event configuration. From our lightcurve observations recorded 12.6 and 15.5 days after the Spitzer/IRS (Fig. 2), we derive an absolute magnitude $H_v(0)$=12.09 ± 0.07 (Table 4), very close to the $H$ value from JPL HORIZONS ($H$=12.10). From NEATM we obtain an effective diameter $D_{eff}$ = 9.7 ± 1.2 km and $p_V$ = 0.27 ± 0.07. The high albedo is in agreement with the taxonomic class of the asteroid. The method described in Behrend et al. (2006) assumed that the two components are prolate ellipsoids of same size orbiting around the center of mass in a circular tidally locked orbit, with identical density. We find the size of the prolate components (with greater diameters $D_a$ = 8.1 ± 1.0 km and $D_b$ = 5.8 ± 0.7 km, see Fig. 5 in Behrend et al. 2006). The bulk density for the system ρ = 0.9 ± 0.2 g/cm$^3$ (Behrend et al. 2006) is surprisingly low, since HED meteorites, the adopted meteorite analogs of V-type asteroids, have bulk densities of 2.86-3.26 g/cm$^3$ (Britt & Consolmagno, 2004; McCausland & Flemming, 2006) implying that, if (854) Frostia has the same composition, its macro-porosity should be between 50 and 80%. (809) Lundia, a T2 V-type binary system with components of the same size as (854) Frostia, has a higher bulk density.

**(3782) Celle:** The binary nature of (3782) Celle was derived by Ryan et al. (2004) from lightcurve observations collected in 2001 to 2003, which revealed a low amplitude (0.10-0.15 mag) variation of $P$ = 3.84 h (the spin of the primary) and anomalous attenuations due to mutual events between the satellite and the primary with a period of 36.57 ± 0.03 h. Ryan et al. (2004) derived a first-order model of this asynchronous binary system (also called T3), assuming a similar albedo and spherical shape for both components and a circular mutual orbit with a period of 36.57 h. Assuming that they detected a total occultation/eclipse, they derived $D_s/D_p$ = 0.43 ± 0.01 and a/$D_p$ = 3.3 ± 0.2. Using NEATM, we estimate $D_{eff}$ = 6.6 ± 0.7 km and $p_V$ = 0.23 ± 0.09. The primary of (3782) Celle has a diameter $D_p$ = 6.1 ± 1.0 km and the satellite, with $D_s$ = 2.6 ± 0.8 km (assumed to be spherical), orbits at 20.1 ± 3.7 km from the primary. From Kepler's law, we derive a total mass $M$ = 3.4 ± 1.5 x $10^{14}$ kg, corresponding to an average density $\rho$ = 2.4 ± 1.1 g/cm$^3$. Considering the bulk density of HED meteorites, we derive a macroporosity between 30 and 40% for this binary asteroid, which is typical for a rubble-pile interior and for most known binary asteroids. Since Nesvorny et al. (2005) determined that (3782) Celle belongs to the Vesta collisional family, it is possible that this binary system is the product of reaccumulated fragments from the south pole impact basin. A subsequent oblique impact could have spun up the loosely-bound asteroid, splitting it in two halves as suggested by Descamps & Marchis (2008).

### 6.4 S-complex asteroids

Because of several space mission visits (NEAR around (433) Eros, Hayabusa hovering above (25143) Itokawa) and flybys (e.g., the Galileo spacecraft flyby of (243) Ida and its companion Dactyl), S-complex asteroids are without any doubt the best characterized asteroids in our Solar System. Samples of (25143) Itokawa returned recently by the Japanese Hayabusa mission confirmed that at least some S-type asteroids are one of the sources of LL and L group equilibrated ordinary chondrites (Yurimoto et al. 2011). The bulk density of Itokawa samples was estimated at 3.4 g/cm$^3$ by Tsuchiyama et al. (2011).

Bus and Binzel (2002) described two parts of the S-complex from visible spectroscopy. The "end members", made of the A, K, L, Q, and R classes, which show reflectance spectra displaying subtle variations in the shape, and "the Core" classes composed of S, Sa, Sk, Sl, Sq, and Sr sub-classes with a similar 1-μm absorption band and UV slope. The addition of NIR reflectance spectroscopy allowed DeMeo et al. (2009) to simplify the S-complex into five sub-classes called S, Sa, Sq, Sr, Sv. Asteroids with a steep red slope receive the notation "w" since this increase of slope arises from space weathering (Clark et al. 2002). When possible, we combined the reflectance spectra in the visible and NIR to determine the taxonomic class from DeMeo et al. (2009). In Table 2, Sw-types (1089) Tama and (1509) Esclangona and S-types (3623) Chaplin and (5407) 1992AX have their taxonomic classes determined from our unpublished observations (Enriquez et al. in preparation 2012). Other asteroid taxonomic classes are based on visible surveys conducted by Bus and Binzel (2002), Lazzaro et al. (2004), Reiss et al. (2009), or near-infrared surveys by Birlan et al. (2011) and Marchis et al. (2011a). In the cases of multiple observations and/or multiple analyses for the same asteroid, the determined taxonomic classes always belong to the S-complex.

**(1089) Tama:** This T2 binary asteroid was the target of 2 large photometric campaigns of observations by Behrend et al. (2006) conducted in the years 2003-2005. The December 2003 lightcurves display the typical U-V shape of similar size binary systems in mutual events. Our lightcurves recorded 18 days before and 32.3 days after our Spitzer/IRS observation show that the binary system was not at the time in a mutual event configuration. The measured absolute magnitude ($Hv(0)$ = 11.63) is very close to the value from JPL HORIZONS ($H$=11.60). Lazzaro et al. (2004) classified (1089) Tama from its reflectance spectrum in the visible in the S-type class. It belongs to the large Flora collisional family (Zappala et al. 1995). Our NEATM analysis allow us to derive an effective diameter $D_{eff}$= 12.2 ± 1.6 km and $p_V$ = 0.27 ± 0.08. Using the model described in Behrend et al, (2006), which assumes that the binary system is made of two prolate ellipsoids components of similar size orbiting the center of mass in a circular, tidally locked orbit (P=0.685 days), with identical density, we derive the size the components (greater diameter of the tidally-locked oblate components $D_a$= 10.4 ± 1.4 km & $D_b$ = 7.1 ± 0.9 km). The bulk density $\rho$ = 2.5 ± 0.4 g/cm$^3$ from Behrend et al. (2006) is similar to the bulk density of (433) Eros ($\rho$ = 2.5 ± 0.8 g/cm$^3$) measured from NEAR spacecraft (Yeomans et al. 1999) and significantly higher than the bulk density of C-complex asteroids. From the bulk density of the (25143) Itokawa sample (3.4 g/cm$^3$), we estimate the macro-porosity to be between 0 and 50%. Based on our visible/NIR reflectance spectrum analysis (Enriquez et al. in preparation 2012), this asteroid belongs to the Sw-class of deMeo et al. (2009) in agreement with its measured high albedo.

**(1313) Berna:** The binary nature of this main-belt asteroid was revealed from lightcurve observations, which showed a U-V shape typical of double systems of similar size (Behrend et al. 2006). A lightcurve with a period of 25.46 h and amplitude of 1.4 mag was derived from a one-month photometric survey in February, 2004. We collected lightcurve observations 49 days after the Spitzer/IRS observation. The lightcurve suggests that the Spitzer spectrum was recorded when the system was at its maximum elongation. From our photometric measurement we derive $Hv(0)$ = 11.69 ± 0.12, which was used in our NEATM model to estimate the effective diameter: $D_{eff}$ = 13.3 ± 1.4 km. Using the model described in Behrend et al. (2006) with prolate ellipsoids of same size and density, orbiting the center of mass in a circular, tidally locked orbit, we can derive the sizes of the tidally-locked oblate components ($D_a$= 10.6 ± 1.1 km and $D_b$ = 8.4 ± 0.9 km) and an average density $\rho$ = 1.3 ± 0.4 g/cm$^3$. This average density is in the upper limit of the one estimated by our colleagues (1.1-1.4 g/cm$^3$) based on a conversion of H into a size estimate for the asteroid. This density is in the upper range of the density for C-complex and P-type asteroids (See sections 6.1 & 6.2), which is surprising since the geometric albedo derived from our NEATM ($p_V$= 0.21 ± 0.09) is close to the measured albedo for S-type asteroids. One possible explanation could be that the assumptions on the shape model of the components made in using the Behrend et al. (2006) are invalid for this binary system. The albedo $p_V$~0.21 confirms that Berna belongs to the S or Sq classes as suggested from NIR observations collected recently (Enriquez et al. in preparation 2012). Also the (1313) Berna emissivity spectrum is similar to (3749) Balam overall, although Balam's spectrum seems to show strong emission features at longer wavelength.

**(1333) Cevenola:** This S-type (Lazzaro et al. 2004) or Sq-type (Birlan et al. 2011) main-belt asteroid belongs to the Eunomia dynamical family (Zappala et al., 1995; Nesvorny et al. 2005). Warner et al. (2002) reported a lightcurve amplitude of 0.97 ±0.03 mag and a period of 4.88 ± 0.02 h from observations recorded on February 4-7, 2002. Since no follow-up observations have been recorded or published, the true nature of (1333) Cevenola is still unknown. We classified it as a T4 binary, since it could be a binary asteroid system or an extremely elongated one made of two joined components, like the main-belt asteroid (3169) Ostro ($D_{eq}$ ~11 km, Descamps et al. 2007) or the large ($D_{eq}$ ~225 km) Trojan asteroid (624) Hektor (Marchis et al., 2006c). From our lightcurve data taken 15 days before and after the Spitzer observations, we measured Hv(0) = 12.05 ± 0.12 (for comparison H = 11.4 in JPL HORIZONS). The amplitude of the lightcurve is ~1.1 mag, so these almost-simultaneous measurements were necessary to derive an accurate H value for our model. Using NEATM, we estimate its size ($D_{eff}$ = 11.2 ± 1.4 km) and its albedo ($p_V$ = 0.21 ± 0.08).

**(1509) Esclangona:** The binary nature of this Hungaria asteroid was revealed by AO observations by Merline et al. (2003). From an analysis of archive images (ESO program ID 70.C-0543) recorded using the VLT-NACO on February 13, 2003, and February 15, 2003, we showed that the satellite has a maximum separation of 0.187 arcsec, corresponding to a projected separation of 132 km. The difference in brightness the two components is $1.4 \pm 0.2$ mag, corresponding to a size ratio $D_s/D_p \sim 0.5 \pm 0.1$, assuming that both components have the same albedo. The angular position of the satellite changed significantly over a period of $\sim 2h$ in observations taken on both dates, implying that the orbit was seen edge-on.

More recently, Warner et al. (2009a) reported the detection of multi-periodicity in the lightcurve of the system, corresponding to the spin of two components. Since no mutual event effects were detected over almost two months of observation, the period of the mutual orbit is unknown. Since the period and the separation remain poorly constrained, the mass is unknown, consequently the density is undetermined. From the NEATM model (Table 5), we derive an effective diameter $D_{eff} = 9.0 \pm 1.0$ km. Figure 2b shows that the two components were not in a mutual event configuration at the time of the Spitzer/IRS observation, since the lightcurve had an amplitude of 0.2 and 0.3 mag, typical of the variation mentioned in Warner et al. (2009a). Consequently, assuming a similar albedo (estimated to $p_V = 0.11 \pm 0.02$) and a spherical shape, the components of (1509) Esclangona have the following diameters: $D_p = 8.0 \pm 0.8$ km and $D_s = 4 \pm 0.7$ km. Based on our unpublished visible/NIR reflectance spectrum analysis this asteroid should belong to the Sw class. Bus and Binzel (2002) classified it, however, in the Ld class.

**(3623) Chaplin:** Like (1333) Cevenola, (3623) Chaplin's lightcurve ($P=8.361 \pm 0.005$ h) is characterized by a large amplitude of $0.97 \pm 0.02$ mag (Birlan et al. 1996), which suggests a bilobated shape or a possible binary nature. Orbital analyses by Zappala et al (1995) and Nesvorny et al. (2005) classified this asteroid in the Koronis dynamical family. Its NIR reflectance spectrum is typical of Sq asteroids (Birlan et al. 2011). Our lightcurves taken 15.8 days before and 14.1 days after our Spitzer/IRS observations show an amplitude of 1.15 mag, indicating that the asteroid is seen edge on. From these photometric observations, we derive a absolute magnitude $Hv(0) = 11.82 \pm 0.10$ at the time of Spitzer/IRS observation, giving a diameter $D_{eff} = 11.1 \pm 1.5$ km and $p_V = 0.27 \pm 0.10$. Additional photometric observations will help to determine the true nature of this asteroid.

**(3749) Balam:** In the family of multiple asteroids, the (3749) Balam system is certainly one of the most mysterious members. Its first satellite was discovered in 2002 using an AO system mounted on the Gemini North telescope. Marchis et al. (2008b) attempted to derive the orbital parameters of this 2-km satellite diameter and found a solution with the satellite orbiting in 60-70 days around the primary in a highly eccentric orbit ($e \sim 0.9$ and $a \sim 290$ km), implying that this binary system belongs to the W-c group defined by Pravec and Harris (2007). More recently, from a careful photometric analysis of the lightcurves of (3749) Balam, Marchis et al. (2008d) suggested that the primary of the system is in fact dual and composed of two components. The NIR reflectance spectrum (Marchis et al. 2011a) showed that this asteroid belongs to the S-complex, and could be a Q- or S- type. Vokrouhlicky (2009) showed that (3749) Balam is a very young system with an age estimated to be less than a million years old, making this asteroid an interesting target to understand the evolution of small Solar System bodies. Because our lightcurve does not show any attenuations at the time of the Spitzer/IRS (from a lightcurve taken 71 days before the Spitzer observations), we assume that the central system was separated, and use the *H, G* values derived by Polishook et al. (2011) to calculate from NEATM an effective diameter $D_{eff}$ = 4.7 ± 0.5 km and $p_V$= 0.27 ±0.10. Assuming that the outer satellite was not occulted at the time of Spitzer/IRS observation and that all the components of the system have the same albedo, we find (using the measured size ratio $D_s/D_p$ = 0.43 ± 0.07 by Marchis et al. (2008b) and assuming that the components are spherical) that the outer moon diameter is $D_s$ = 1.8 ± 0.6 km, and the two central components have diameters $D_{p1}$ = 4.3 ± 0.5 km and $D_{p2}$= 1.7 ± 0.6 km. Vachier et al. (2011) determined a bundle of appropriate solutions for (3749) Balam's outer satellite with *a* varying from 189 to 298 km, *P* from 54 to 162 days and *e* from 0.35 to 0.77, leading to a density from 1.7 to 3.7 g/cm³, typical for S-type asteroids (Marchis et al. 2011a). A unique solution could be derived from additional future observations taken with the Hubble Space Telescope or large 8-10m class telescopes equipped with AO. (3749) Balam is an interesting system since it is could be an easy target for a space mission dedicated to exploring young and multiple asteroid systems (e.g., Diversity mission concept by Marchis et al. (2011b)).

**(4674) Pauling:** The existence of the satellite of (4674) Pauling was revealed by Merline et al. (2004). Seven years later, its orbit remained unknown and not much else is known about this binary system. After reanalyzing archive VLT/NACO observations (ESO program ID 72.C-0753) taken on March 4, 2004, from 5:29 UT to 7:10 UT, we determine an angular separation varying from 0.364 arcsec to 0.354 arcsec (corresponding to a projected distance of 233 ± 4 km) and a size ratio of 0.349 ± 0.004 between the primary and the secondary measured in H and K bands (~1.6 and ~2.2 μm). We used an additional observation on May 28 2007 at 10:15 UT with the W.M. Keck II AO and its NIRC2 camera in the K band to derive a similar size ratio (~0.4) and an angular separation of 0.416 arcsec, corresponding to a projected distance of ~300 km. From a reflectance spectrum collected in the NIR using IRTF, Marchis et al. (2011a) showed that this asteroid belongs to the Q- or Sq- subclasses. The lightcurve of (4674) Pauling recorded only 19 days before the Spitzer/IRS observation, shows no large flux variation, implying that the primary has a regular shape. Using the absolute magnitude from JPL HORIZONS (H=13.3) and the standard deviation error on the lightcurve at the time of Spitzer/IRS observation (0.03) as the uncertainty on H, we derive using NEATM an effective diameter $D_{eff}$ = 4.7 ± 0.5 km and $p_V$= 0.39 ± 0.09. Considering that both components have the same albedo, (a likely assumption since the size ratio does not change in H and K bands; we estimate the size of the primary and satellite to be $D_p$ = 4.4 ± 0.5 km and $D_s$ = 1.6 ± 0.5 km (assuming that the components are spherical). (4674) Pauling is interesting since it is one of the few resolved loosely-bound asteroid systems (W-group by Pravec & Harris, 2007) with $a/R_p \sim$ 89, assuming an eccentricity for the mutual orbit *e*~0.35, like (3749) Balam. This asteroid has the highest geometric albedo in our sample of Q- or S-type asteroid (Marchis et al. 2011a), which may imply a composition or a surface age different from other S-types (average $p_V$ = 0.26 ± 0.03). No asteroid pairs or collisional families seem to be linked with (4674) Pauling, making this a unique loosely-bound asteroid.

**(5407) 1992 AX:** This Sk-type (Bus and Binzel, 2002) Mars-crossing asteroid is listed as a probable binary by Pravec et al. (2000), who analyzed photometric observations taken in January and February, 1997. The lightcurve can be decomposed into a short-period ($P_1$~2.54 h) component and long-period ($P_2$~6.76 h) component with amplitudes of 0.11 and 0.08 mag, respectively. Based on the profile of the attenuations, it is likely that the long-period component is caused by mutual events, plus a contribution from the secondary moon that is tidally locked.

They derived a lower limit of 0.30 for $D_s/D_p$. No lightcurves were recorded during the Spitzer/IRS observations, consequently we use the absolute magnitude $H$=14.47 and an assumed error of 0.10 to calculate the equivalent diameter and geometric albedo from NEATM. We obtain $D_{eff}$ = 3.8 ± 0.4 and $p_V$ = 0.20 ± 0.08. With $D_s/D_p$ = 0.30, the primary and satellite diameters are 3.6 ± 0.4 km and 1.1 ± 0.4 km respectively assuming that they are circular.

**(5905) Johnson:** Since the discovery of its binary nature (Warner et al. 2005b), this Hungaria Q/Sq-type (Enriquez et al. in preparation 2012) asteroid received a lot of attention from the community of observers and several new lightcurves were recorded again from May to June, 2008. From this new dataset, Warner et al. (2009b) showed that the lightcurve can be decomposed into a short-period component with a period of 3.78 h and amplitude of 0.08, corresponding to the spin of the primary, superimposed on a long-period component of 21.78 h, with an amplitude of 0.15-0.22, corresponding to mutual events between the two components. A precise size ratio was derived from that study: $D_s/D_p$ = 0.38 ± 0.02. Our lightcurve observation recorded 0.43 days (almost one long-period) after the Spitzer/IRS observation (Fig. 2d), has an amplitude of 0.25 mag, suggesting that the IRS spectrum was recorded when the system was in a mutual event configuration. Based on this unique low SNR lightcurve, it is impossible to detect if the system was observed in transit and/or in eclipse. A complete analysis of this system combining our mid-IR data with a long-term photometric survey could help deriving the characteristics of the system (size & shape of the components, mutual orbit and density).

We used the absolute magnitude ($H$=14.0 ± 0.1) from JPL HORIZONS, which is identical to the Warner et al. (2009b) measurements, to derive an effective diameter $D_{eff}$ = 4.1 ± 0.5 km and $p_V$ = 0.27 ± 0.10. This relatively high albedo is in agreement with the NIR reflectance spectrum recorded with IRTF on July, 2011, which suggests that (5905) Johnson is a Q- or Sq- type asteroid.

This system may be even more complicated since Warner et al. (2009b) reported a third period in their photometric survey that they interpreted as the rotation of the satellite, or a third body in the system.

**(69230) Hermes:** Shortly after this Apollo asteroid, previously known as 1937UB, was rediscovered by Skiff et al. (2003) from LONEOS observations, its binary nature was revealed by radar observations (Margot et al. 2003). The data showed that (69230) Hermes is made of two similar size components (binary class T2) with diameters of 300-450 m, orbiting their center of mass in less than 13-21 h. We did not collect any lightcurves of this asteroid, so we used the JPL HORIZONS absolute magnitude $H$=17.48 ± 0.10. We obtain an effective diameter $D_{eq}$ = 0.8 ± 0.1 km and $p_V$ = 0.265 ± 0.10, in agreement with its S-type classification derived from NIR spectrum analysis by Rivkin et al. (2004). Our thermal analysis suggests that the two components have an effective diameter of 566 ± 71 m and an albedo $p_V$=0.26 ± 0.010. The diameter is in the upper range of estimated size from radar observations briefly described in Margot et al. (2003). (69230) Hermes is the smallest asteroid observed in our sample of 28 binary asteroid systems.

**(329437) 1994 AW1:** Mottola et al. (1995) reported a complex lightcurve for this S-type (Bus and Binzel, 2002) Amor asteroid with two periods. Follow-up observations and their analysis described in Pravec & Hahn. (1997) suggested that 1994AW1 could be a binary asteroid system with a long period component of 22.3 h and a short one of 2.51 h. As in the case of (5905) Johnson described above, a precise size ratio was derived from that study: $D_s/D_p$ = 0.53± 0.02.

We did not collect a lightcurve at the same time as the Spitzer/IRS observations, so we use the absolute magnitude from JPL HORIZONS ($H$=16.97 with an estimated error of 0.1). The effective diameter ($D_{eff}$ = 1.0 +0.1) and geometric albedo ($p_V$ = 0.28 ± 0.10) were derived from our NEATM analysis. The diameters of the components of 1994AW1 are estimated to 883 ± 10 m and 468 ± 10 m for the primary and secondary respectively, assuming that the system was not observed in a mutual event configuration and are spherical in shape.

## 6.5 Asteroid systems with unknown taxonomic class

Three asteroids from our sample do not yet have a classification from reflectance spectroscopy in the visible and/or NIR. Two of them ((4492) Debussy and 2000 RG79) are part of the asteroid main belt and the third one (2000 DP107) is an Apollo. The determination of the sizes and densities of the system components does not depend on the knowledge of their taxonomic class.

**(4492) Debussy:** Similarly to (1313) Berna, the binary nature of (4492) Debussy, a main-belt asteroid, was suggested from the U-V shape and large amplitude (A~1.2, P=26.58 h) of its lightcurve recorded over three epochs of observations from Oct., 2002, and April, 2005 (Behrend et al. 2006). The lightcurve recorded 19, 20, and 39 days after our Spitzer/IRS observations shown in Fig. 2a, suggest that the observation was taken when both components were separated. We derived a absolute magnitude $Hv(0)$ = 13.05 ± 0.07 from our photometric measurement, which was used in our NEATM model to derive a diameter $D_{eff}$ = 16.5 ± 1.9 km. Once again we can use the model described in Behrend et al (2006) to derive the greater diameter of the prolate components ($D_a$= 14.6 ± 1.7 km & $D_b$ = 9.4 ± 1.1 km). The density 0.9 ± 0.1 g/cm$^3$ derived in Behrend et al. (2006) is very close to those derived for C-complex and P-type asteroids in sections 6.1 & 6.2. Additionally, the geometric albedo derived with our NEATM model ($p_V$ = 0.04 ± 0.02) is very close to that of C-complex and P-type asteroids. Since the (4492) Debussy emissivity spectrum partially mimics the emissivity bands of (45) Eugenia (see Section 5.2.1), it is very likely that this asteroid is a member of the C-complex.

**(76818) 2000RG79:** This asteroid was found to be a binary by Warner et al. (2005a) and was surveyed photometrically from October, 2008, to January, 2009. They found a primary period of $P_1$ = 3.116 h with an amplitude $A_1$ = 0.14 mag (the spin of the large component) and an orbital period of 14.12 h with amplitude varying from 0.12 to 0.15 mag (Warner and Stephens 2009c). From these data, our colleagues estimated a size ratio $D_s/D_p$ = 0.32 ± 0.03. We did not record lightcurve observations during our Spitzer/IRS observation, so we used the absolute magnitude H=14.26 ± 0.10 from JPL HORIZONS. The effective diameter ($D_{eff}$= 2.6 ± 0.3 km) was derived from our NEATM analysis. The diameters of the components of 2000RG79 are estimated to 2.5 ± 0.3 km and 0.8 ± 0.1 km for the primary and secondary respectively, assuming that the system was not observed in a mutual event configuration. The extremely high geometric albedo of 0.50 ± 0.15, the highest measured in our survey, suggests that i) this asteroid is a member of the S-complex (possibly a V-type asteroid, since (4) Vesta has an albedo of ~0.4 (see Tedesco et al. 2002) or E-type , ii) its reported absolute magnitude is incorrect.

**(185851) 2000DP107:** The binary nature of this asteroid was announced by Ostro et al. (2000), who discovered the secondary using Goldstone radar observations. Follow-up Arecibo radar observations recorded in October, 2000, gave estimated diameters for the primary and secondary of 800 m and 300m, respectively (Margot et al. 2002), with an orbital period of 1.755 days and a semi-major axis of 2.620 km, implying a mass of $4.6 \pm 0.5 \times 10^{11}$ kg. Lightcurve analysis through 3D-shape modeling by Scheirich and Pravec (2009) provides a $D_s/D_p$ ratio of 0.43. $_{0.04}^{+0.3}$. For our NEATM analysis, we were unable record a lightcurve at the time of Spitzer/IRS observation, so we used H= 18.03±0.10 from JPL HORIZONS. We derive a diameters $D_p$ = 890 ± 120 m and $D_s$= 450 ± 160 m (3-sigma error), in agreement with the Arecibo radar observations. We derived an average density 0.9 ± 0.3 g/cm$^3$. Our density measurement is close to the density of C-complex & P-type asteroids. Our NEATM analysis gave $p_V$ = 0.11 ±0.04, suggesting that 2000DP107 could be indeed a C-complex asteroid.

## 7. Conclusion

This work describes the data processing and analysis of mid-infrared spectra of 28 multiple asteroid systems that are part of the main-belt and the NEA populations representing samples of various sub-groups of binaries, taxonomic classes, and with a wide range of sizes (with $D$ from 0.8 km to 300 km).

Considering their taxonomic classes, we find that:

- C-complex binary asteroids in T1 binary systems have a low bulk density between 0.7 and 1.7 g/cm$^3$. The G-type (130) Elektra seems to be the densest C-complex asteroid with a density similar to (93) Minerva and (1) Ceres (Marchis et al., 2011c).
- P-type binary asteroids in T1-binary systems also have low bulk densities, varying from 0.7 to 1.7 g/cm$^3$, similar to the density of the Trojan (617) Patroclus (Mueller et al. 2009).
- Our independent TPM analysis confirms the high density of the M-type asteroid (22) Kalliope already suggested in Descamps et al. (2008), and which is similar to the other M-types (216) Kleopatra (Descamps et al. 2011) and the planetesimal (21) Lutetia (Vernazza et al. 2011).
- (9069) Hovland ($D_{eff}$ ~3 km, $p_V$~0.4) could be the only E-type asteroid in our sample, but its emissivity spectrum cannot be used to characterize the possible meteorite analog due to its low SNR.
- We have three asteroids classified as V-type in our sample; none of them has an albedo as high as (4) Vesta and their densities are significantly less than the bulk density of HED meteorites, implying a significant macroporosity or contamination by less dense material.
- The fourteen S-complex asteroids in our sample are diverse in size ($D_{eff}$ from 1 to 13 km) but all of them are T2, T3 or T4 asteroids. The bulk densities of the two S-type asteroids with known mutual orbits range from 1.3 to 3.7 g/cm$^3$.

From the geometric albedos and the profiles of their emissivity spectra, we were able to infer the taxonomic complex of two unclassified asteroids. (4492) Debussy has a low albedo ($p_V$ ~ 0.04) and a similar emissivity spectrum to (45) Eugenia, so it is most likely a C-complex asteroid. (185851) 2000DP107 has the expected low density and albedo of a C-complex body, but a similar emissivity profile to 1999HF1, an X-type asteroid.

The emissivity spectra of asteroids larger than 130 km show obvious reststrahlen bands and the Christiansen peak at wavelength <12 μm, implying that their surfaces are composed of fine-grained regoliths characteristic of a mature surface. Combining visible/NIR reflectance spectra with the emissivity spectra of these eight T1 asteroids could help determine their surface compositions. Our TPM analysis suggests that these asteroid surfaces have a low thermal inertia ($\Gamma \leq \sim 100$ J s$^{-1/2}$K$^{-1}$m$^{-2}$), confirming that they are covered with a layer of thermally insulating material. This thick layer of regolith, formed by micro-meteorite bombardment can remain on the large asteroids because of their significant gravity.

There is a gap in our target sample, since it does not contain any asteroids with a diameter between 17 km and 130 km. This might be due to a combination of observational bias, since small ($D_s < 1/10$ x $D_{eq}$) companions of 15-80 km asteroids located in the main belt (small T1) cannot be detected using high angular resolution imaging provided by current AO systems and HST, and a fundamental difference in binary formation mechanism between the 1- to 10-km-scale primaries and the 100-km-scale primaries. It is likely that since the formation of T3 multiple asteroids is mostly driven by the YORP effect, large ($D_{eff} > 17$ km) T2 and T4 asteroids should be extremely rare.

The smaller ($D_{eq} < 17$ km) asteroids of our sample encompasses T3, T2 and T4 binary groups that may be the products of the evolutionary track of a rubble-pile asteroid (Jacobson & Scheeres, 2011a) and also the W group that includes well separated, small binary systems (Pravec and Harris, 2007) whose formation is not well understood. Their emissivity spectra show emissivity lines, but they are of significantly lower contrast, implying that the regolith is coarser than the large T1 binary asteroids. It is unclear if a recent formation, a weaker gravity, or a different composition can explain the lower spectral contrast of emissivity features for several small asteroids.

All of the data used in this investigation, including the SED for each asteroid and the emissivity spectra rebinned to $R$=60 at 20 µm are made available in the electronic supplement, Table S2. We hope to encourage further analysis of these data by comparison with lab spectra of meteorites and minerals considering different grain size distributions and composition to refine the surface composition of these asteroids. Similar works were performed by our colleagues for Trojan asteroids (Emery et al. 2006), the V-type (956) Elisa (Lim et al. 2011), and Vernazza et al (2011) for large asteroids of our sample.

Our sample presented here contains a limited sample of 28 of the more than 195 multiple asteroid systems known. The mid-IR observations are essential for the determination of sizes, densities and surface compositions of these bodies. A continuation program with Spitzer/IRS is not possible since the instrument was decommissioned when the Spitzer telescope lost its cryogenic capability. Future low-resolution spectroscopic capabilities of multiple asteroids depend on the development of instruments such as SOFIA/FORCAST, a balloon-based observatory, or future mid-IR space telescopes such as JWST.

**Acknowledgement:** The authors would like to thank Prof. Alan Harris from DLR and an anonymous referee for their valuable comments which improved significantly the quality of this article.This work is based in part on observations made with the Spitzer Space Telescope, which is operated by the Jet Propulsion Laboratory, California Institute of Technology under a contract with NASA. Support for this work was provided by NASA through an award issued by JPL/Caltech. FMA, JEN & MBA were supported by the National Science Foundation under award number AAG-0807468.

**TABLES**

**Table 1:** Observing circumstances for the Spitzer/IRS data. *Tint & coadds* indicate the integration time per observations and the number of coadds. *Tint × coadds* is the total exposure time per observing mode (SL2, SL1, LL2,LL1,SH or LH, see Section 2.2). The column named $\Delta_{phase}$ corresponds to the difference in solar phase at the time of our Spitzer observation and our groundbased observations.

| ID | Provisional Designation (or name) | Date and time of Observations HH:MM (+/-MM) | Integration times SL2,SL1,LL2,LL1,SH,LH (format is:  tint-coadds) | $\Delta_{AU}$ distance Spitzer (AU) | $r_{AU}$ distance Sun (AU) | phase angle (deg) | $\Delta_{phase}$ deg | error D in % | error pV in % |
|---|---|---|---|---|---|---|---|---|---|
| 45 | Eugenia | 2007-10-30 09:05 +/- 5 | 6-2,6-2,---,---,6-2,6-2 | 2.93 | 2.61 | 20.2 | 5.4 | 9 | 17 |
| 130 | Elektra | 2005-04-22 05:53+/-5 | 14-1,6-1,---,---,6-1,6-1 | 3.50 | 2.91 | 14.6 | 1.4 | 2 | 4 |
| 762 | Pulcova | 2004-09-28 14:40 +/-20 | 14-2,14-2,---,---,6-2,6-2 | 2.98 | 2.90 | 20.0 | 0.7 | 1 | 2 |
| 22 | Kalliope | 2005-11-20 05:25+/-5 | 6-2,6-2,---,---,6-2,6-2 | 2.77 | 2.27 | 20.3 | 0.5 | 1 | 2 |
| 87 | Sylvia | 2004-11-12 05:50 +/-05 | 6-1,6-1,---,---,6-1,6-1 | 3.22 | 2.71 | 17.1 | 0.6 | 1 | 2 |
| 107 | Camilla | 2008-02-29 20:00 +/- 5 | 6-2,6-2,---,---,6-2,6-2 | 3.39 | 3.19 | 17.2 | 4.2 | 7 | 14 |
| 121 | Hermione | 2007-10-01 00:03 +/- 5 | 6-2,6-2,---,---,6-2,6-2 | 3.36 | 2.94 | 17.1 | 0.0 | 0 | 0 |
| 216 | Kleopatra | 2006-02--2 06:14 +/-7 | 60-1,6-2,---,---,6-2,6-2 | 3.15 | 3.35 | 17.5 | 2.4 | 4 | 8 |
| 283 | Emma | 2008-05-27 05:03 +/- 10 | 6-2,6-2,---,---,6-2,6-2 | 2.92 | 2.61 | 20.0 | 6.5 | 10 | 21 |
| 9069 | Hovland | 2008-01-10 12:01 +/- 59 | 60-5,60-3,120-4,120-4,---,--- | 1.80 | 1.36 | 33.9 | 11.6 | 19 | 37 |
| 137170 | 1999 HF1 | 2008-05-31 09:35 +/- 10 | 14-4,6-3,14-2,30-2,---,--- | 1.16 | 0.71 | 59.4 | 1.0 | 2 | 3 |
| 809 | Lundia | 2008-05-27 16:37 +/- 29 | 60-3,14-3,30-2,30-2,---,--- | 2.11 | 1.83 | 28.6 | 7.2 | 12 | 23 |
| 854 | Frostia | 2008-08-12 18:55 +/- 20 | 60-4,14-2,30-2,30-2,---,--- | 2.24 | 2.09 | 26.9 | 4.0 | 6 | 13 |
| 3782 | Celle | 2008-05-27 17:34 +/- 62 | 60-5,60-3,120-4,120-5,---,--- | 2.21 | 1.97 | 27.1 | 7.1 | 11 | 23 |
| 1089 | Tama | 2008-03-30 04:48 +/- 25 | 60-5,14-5,30-3,30-3,---,--- | 2.42 | 1.97 | 23.7 | 11.7 | 19 | 38 |
| 1313 | Berna | 2007-10-06 T05:04 +/- 17 | 60-3,14-2,14-2,14-4,---,--- | 2.11 | 1.71 | 29.0 | 7.2 | 12 | 23 |
| 1333 | Cevenola | 2008-05-27 15:47 +/- 37 | 60-5,60-2,120-2,120-2,---,--- | 2.98 | 2.71 | 19.7 | 6.0 | 10 | 19 |
| 1509 | Esclangona | 2008-01-10 11:10 +/- 20 | 60-3,14-3,30-4,30-3,---,--- | 1.84 | 1.45 | 33.3 | 11.6 | 19 | 37 |
| 3623 | Chaplin | 2008-02-25 07:12 +/- 59 | ---,60-8,120-4,120-4,---,--- | 3.08 | 2.95 | 19.0 | 3.6 | 6 | 12 |
| 3749 | Balam | 2008-01-15 01:35 +/- 59 | 60-5,60-3,120-3,120-5,---,--- | 2.15 | 1.63 | 26.9 | 0.9 | 1 | 3 |
| 4674 | Pauling | 2007-10-03 22:11 +/- 42 | 60-5,60-2,120-2,120-3,---,--- | 1.51 | 1.99 | 30.3 | 0.9 | 1 | 3 |
| 5407 | 1992 AX | 2007-08-31 22:47 +/- 08 | 60-2,14-2,30-2,30-4,---,--- | 1.39 | 1.04 | 47.2 | 6.0 | 10 | 19 |
| 5905 | Johnson | 2008-05-05 18:49 +/- 36 | 60-5,14-5,120-2,120-2,---,--- | 1.93 | 1.72 | 31.4 | 2.5 | 4 | 8 |
| 69230 | Hermes | 2008-05-27 14:53 +/- 26 | 60-2,14-2,30-4,120-3,---,--- | 1.13 | 0.28 | 57.3 | 5.0 | 8 | 16 |
| 329437 | 1994 AW1 | 2008-04-21 T06:27 +/- 10 | 60-5,60-3,120-4,120-5,---,--- | 1.15 | 0.67 | 60.7 | 0.4 | 1 | 1 |
| 4492 | Debussy | 2007-11-04 11:43 +/- 49 | 60-5,60-3,120-3,120-3,---,--- | 2.35 | 2.16 | 25.8 | 3.8 | 6 | 12 |
| 76818 | 2000 RG79 | 2007-08-28 21:03 +/- 54 | 60-5,60-3,120-4,120-3,---,--- | 2.11 | 1.71 | 28.9 | 1.6 | 3 | 5 |
| 185851 | 2000 DP107 | 2008-06-01 08:47 +/- 06 | 14-2,6-2,6-2,14-2,---,--- | 1.03 | 0.09 | 72.6 | 2.6 | 4 | 8 |

| (1) | S3OS2 BB, Lazzaro et al. (2004) |
|---|---|
| (2) | Bus and Binzel (2002) |
| (3) | Rivkin et al. (2004) |
| (4) | Tholen and Barucci (1989) |
| (5) | Reiss et al 2009 |
| (6) | Pravec et al. 2002 |
| (7) | Pravec et al. 2000 |
| (8) | Marchis et al., 2011 |
| (9) | Enriquez et al. 2012 |
| (10) | Birlan et al. 2011 |

| ID | Provisional Designation (or name) | H Mag | G | Population skybot | assumed Albedo (pv) | Estimated Diam. (km) | IRAS Diam | size ratio Ds/Dp | Binary Type | Taxonomy | Reference of taxonomy |
|---|---|---|---|---|---|---|---|---|---|---|---|
| 45 | Eugenia | 7.46 | 0.07 | MB IIb | 0.04 | 214 | 215.0 | 0.03 | T1-2moons | FC+C,C | (2), (9) |
| 130 | Elektra | 7.12 | 0.15 | MB IIIb | 0.08 | 182 | 182.0 | 0.04 | T1 | G,Ch,Cgh | (4),(2),(9) |
| 762 | Pulcova | 8.28 | 0.15 | MB IIIb | 0.05 | 137 | 137.0 | 0.14 | T1 | C+Cb,F,Ch | (1),(4),(9) |
| 22 | Kalliope | 6.45 | 0.21 | MB IIIa | 0.14 | 181 | 181.0 | 0.15 | T1 | M+X,X | (2),(5) |
| 87 | Sylvia | 6.94 | 0.15 | MB | 0.04 | 261 | 261.0 | 0.06 | T1-2moons | X+X,P+X | (1),(2) |
| 107 | Camilla | 7.08 | 0.08 | MB | 0.05 | 222 | 223.0 | 0.06 | T1 | X+X,C+X,L | (1),(2),(9) |
| 121 | Hermione | 7.31 | 0.15 | MB | 0.05 | 209 | 209.0 | 0.06 | T1 | C+Ch,Xk/Cgh/K | (2),(8) |
| 216 | Kleopatra | 7.30 | 0.29 | MB | 0.12 | 133 | 125.0 | 0.06 | T1-2moons | Xe | (2) |
| 283 | Emma | 8.72 | 0.15 | MB IIIb | 0.03 | 149 | 148.0 | 0.06 | T1 | C+C,X,X | (1),(4),(9) |
| 9069 | Hovland | 14.40 | 0.15 | MB | 0.23 | 4 | | unk | T3 | Xe | (5) |
| 137170 | 1999 HF1 | 14.41 | 0.15 | Aten | 0.23 | 4 | | 0.23 | T3 | E/M/P,L/Xe/Xc/Ch | (6),(8) |
| 809 | Lundia | 11.80 | 0.15 | MB | 0.23 | 12 | | 0.85 | T2 | V+V,V | (1), (9) |
| 854 | Frostia | 12.10 | 0.15 | MB I | 0.23 | 11 | | 0.89 | T2 | V | (10) |
| 3782 | Celle | 12.50 | 0.15 | MB I | 0.43 | 6 | | 0.43 | T3 | V | (2) |
| 1089 | Tama | 11.60 | 0.15 | MB | 0.24 | 13 | 12.9 | 0.90 | T2 | S+S,Q/Sq, Sw | (1),(8), (9) |
| 1313 | Berna | 11.80 | 0.15 | MB II | 0.23 | 12 | | 1.00 | T2 | Q/Sq | (9) |
| 1333 | Cevenola | 12.10 | 0.15 | MB II | 0.23 | 15 | | 1.00 | T4 | S+Sq,Sq | (1),(10) |
| 1509 | Esclangona | 12.64 | 0.15 | MB | 0.23 | 8 | 8.2 | 0.33 | T3 | A+Ld,Ld,Sw | (1),(5),(9) |
| 3623 | Chaplin | 12.10 | 0.15 | MB IIIa | 0.23 | 11 | | 1.00 | T4 | L,S,Sq | (5),(9),(10) |
| 3749 | Balam | 13.40 | 0.40 | MB | 0.23 | 6 | | 0.43 | W | Q/Sq | (8) |
| 4674 | Pauling | 13.30 | 0.15 | MB | 0.23 | 6 | | 0.32 | W | Q/Sq | (8) |
| 5407 | 1992 AX | 14.47 | 0.15 | Amor II | 0.23 | 4 | | 0.20 | T3 | Sk, (S), S | (2),(7),(9) |
| 5905 | Johnson | 14.00 | 0.15 | MB | 0.23 | 4 | | 0.39 | T3 | Q/Sq | (12) |
| 69230 | Hermes | 17.48 | 0.15 | Apollo | 0.23 | 1 | | 0.90 | T2 | S | (3) |
| | 1994 AW1 | 16.97 | 0.15 | Amor I | 0.23 | 1 | | 0.48 | T3 | Sa | (2) |
| 4492 | Debussy | 12.90 | 0.15 | MB IIb | 0.23 | 7 | | 0.91 | T2 | unknown | |
| 76818 | 2000 RG79 | 14.26 | 0.45 | MB | 0.23 | 4 | | 0.35 | T3 | unknown | |
| 185851 | 2000 DP107 | 18.03 | 0.15 | Apollo | 0.23 | 1 | | 0.41 | T3 | C? | (6) |

**Table 3:** Photometric observations collected using various telescopes used to calculate the absolute magnitude $H_v(0)$ of several asteroids at the time of Spitzer observations. To measure the absolute magnitude we averaged several photometric measurements (shown in Fig. 2) in the fraction of primary rotation range ($\Delta\varphi$) indicated in column 6. When the data were taken in I or R bands we applied the color correction listed in column 5 to derive the absolute magnitude in V band, considering the color indices of the sun (Hardorp, 1980).

| Target | LC Date | Telescope | Hv(0) mag | Assumed Color | phase range Δφ | P hr |
|--------|---------|-----------|-----------|---------------|----------------|------|
| Celle | 2008-06-07 | 1.6m B&C at Pico dos Dias | 13.23 | V-I= 0.88 | 0.05 | 3.84 |
| | 2008-06-09 | 1.6m B&C at Pico dos Dias | 13.01 | V-I =0.88 | 0.05 | |
| Debussy | 2007-11-23 | 0.6m Prompt4 at Cerro Tololo | 13.08 | V-R = 0.54 | 0.05 | 26.606 |
| | 2007-11-24 | 1m Nickel at Lick Observatory | 13.08 | V-R = 0.54 | 0.05 | |
| | 2007-12-13 | 1m Nickel at Lick Observatory | 12.98 | V-R = 0.54 | 0.05 | |
| 1999_HF1 | 2008-03-05 | 0.6m Prompt4 at Cerro Tololo | | | 0.50 | 2.319 |
| | 2008-03-06 | 0.6m Prompt4 at Cerro Tololo | | | 0.50 | |
| Esclangona | 2007-12-02 | 0.6m Prompt4 at Cerro Tololo | 13.28 | V-R = 0.54 | 0.05 | 3.247 |
| | 2008-03-05 | 0.6m Prompt4 at Cerro Tololo | 13.28 | V-R = 0.54 | 0.05 | |
| Berna | 2007-11-24 | 0.6m Prompt5 at Cerro Tololo | 11.57 | V-R = 0.54 | 0.10 | 25.46 |
| | 2007-11-25 | 0.6m Prompt5 at Cerro Tololo | 11.80 | V-R = 0.54 | 0.10 | |
| Cevenola | 2008-06-14 | 0.6m Prompt3 at Cerro Tololo | 12.28 | V-R = 0.54 | 0.02 | 4.88 |
| | 2008-05-13 | 0.6m Prompt4 at Cerro Tololo | 11.95 | V-R = 0.54 | 0.02 | |
| | 2008-05-16 | 0.6m Prompt4 at Cerro Tololo | 11.91 | V-R = 0.54 | 0.02 | |
| Chaplin | 2008-03-10 | 1m Nickel at Lick Observatory | 11.72 | V-R = 0.54 | 0.01 | 8.361 |
| | 2008-02-10 | 0.6m Prompt4 at Cerro Tololo | 11.93 | V-R = 0.54 | 0.01 | |
| Frostia | 2008-08-28 | 1m Nickel at Lick Observatory | 12.03 | V-R = 0.54 | 0.05 | 37.56 |
| | 2008-08-25 | 1m Nickel at Lick Observatory | 12.16 | V-R = 0.54 | 0.05 | |
| Hovland | 2008-01-06 | 0.6m Prompt4 at Cerro Tololo | 14.40 | V-R = 0.54 | 0.05 | 4.217 |
| Pauling | 2007-09-14 | 1.6m B&C at Pico dos Dias | | | 0.30 | 2.53 |
| Tama | 2008-03-12a | 0.6m Prompt4 at Cerro Tololo | 11.63 | V-R = 0.54 | 0.05 | 16.44 |
| | 2008-03-12b | 1m Nickel at Lick Observatory | 11.62 | V-R = 0.54 | 0.05 | |
| | 2008-05-01 | 0.6m Prompt4 at Cerro Tololo | 11.62 | V-R = 0.54 | 0.05 | |
| Johnson | 2008-05-06 | 0.6m Prompt at Cerro Tololo | | | 0.50 | 3.782 |
| Balam | 2007-11-04 | 0.6m Prompt4 at Cerro Tololo | | | 0.50 | 2.805 |
| Lundia | 2008-05-08 | 0.6m Prompt4 at Cerro Tololo | 12.82 | V-R = 0.54 | 0.05 | 15.4 |

**Table 4:** Absolute magnitudes ($H_v(0)$) and errors derived from our lightcurve data or extracted from JPL-HORIZONS (Hv) that will be used in the NEATM. We used the absolute magnitude and G parameter of (3749) Balam from Polishook et al. (2011).

| ID | Name | measured Hv(0) | error Hv(0) | Hv HORIZONS | error Hv | G | error G |
|---|---|---|---|---|---|---|---|
| 45 | Eugenia | | | 7.46 | 0.1 | 0.15 | 0.00 |
| 121 | Hermione | | | 7.31 | 0.1 | 0.15 | 0.00 |
| 107 | Camilla | | | 7.08 | 0.1 | 0.15 | 0.00 |
| 283 | Emma | | | 8.72 | 0.1 | 0.15 | 0.00 |
| 87 | Sylvia | | | 6.94 | 0.1 | 0.15 | 0.00 |
| 762 | Pulcova | | | 8.28 | 0.1 | 0.15 | 0.00 |
| 130 | Elektra | | | 7.12 | 0.1 | 0.15 | 0.00 |
| 22 | Kalliope | | | 6.45 | 0.1 | 0.21 | 0.00 |
| 216 | Kleopatra | | | 7.30 | 0.1 | 0.29 | 0.00 |
| 809 | Lundia | 12.82 | 0.02 | 11.80 | 0.1 | 0.05 | 0.00 |
| 854 | Frostia | 12.09 | 0.07 | 12.10 | 0.1 | 0.35 | 0.00 |
| 1333 | Cevenola | 12.05 | 0.12 | 11.40 | 0.1 | 0.05 | 0.00 |
| 3623 | Chaplin | 11.82 | 0.10 | 12.10 | 0.1 | 0.15 | 0.00 |
| 3749 | Balam[1] | | 0.09 | 13.66 | 0.03 | 0.40 | 0.02 |
| 1089 | Tama | 11.63 | 0.00 | 11.60 | 0.1 | 0.08 | 0.00 |
| 1313 | Berna | 11.69 | 0.12 | 11.80 | 0.1 | 0.15 | 0.00 |
| 1509 | Esclangona | 13.28 | 0.00 | 12.64 | 0.1 | 0.40 | 0.00 |
| 3782 | Celle | 13.12 | 0.12 | 12.50 | 0.1 | 0.05 | 0.00 |
| 4492 | Debussy | 13.05 | 0.07 | 12.90 | 0.1 | 0.05 | 0.00 |
| 4674 | Pauling | | 0.04 | 13.30 | 0.1 | 0.15 | 0.00 |
| 5407 | 1992_AX | | | 14.47 | 0.1 | 0.15 | 0.00 |
| 5905 | Johnson | | 0.12 | 14.00 | 0.1 | 0.15 | 0.00 |
| 9069 | Hovland | 14.40 | 0.03 | 14.40 | 0.1 | 0.10 | 0.00 |
| 69230 | Hermes | | | 17.48 | 0.1 | 0.15 | 0.00 |
| 76818 | 2000 RG79 | | | 14.26 | 0.1 | 0.45 | 0.00 |
| 137170 | 1999 HF1 | | 0.11 | 14.41 | 0.1 | 0.15 | 0.00 |
| | 1994 AW1 | | | 16.97 | 0.1 | 0.15 | 0.00 |
| 185851 | 2000 DP107 | | | 18.03 | 0.1 | 0.15 | 0.00 |

**Table 5:** Characteristics of the asteroids derived from the modeling of their SEDs using NEATM model (see Section 3). The densities and 1-sigma errors of 16 confirmed binary asteroids were calculated using the mass of the system derived from the mutual orbits of the satellite(s).

1. Marchis et al. (2008a)
2. Descamps et al. (2009)
3. Marchis et al. (2008b)
4. Vachier et al. (2011)
5. Marchis et al. (2005)
6. Descamps et al. (2011)
7. Behrend et al. (2006)
8. Ryan et al. (2004)
9. Vachier et al. (2012)
10. Margot et al. (2002)

| Asteroid ID | Provisional Designation (or name) | NEATM model Deff | 3-sigma (Deff) | p_v | 3-sigma (pv) | beaming f eta | 3-sigma (eta) | Mass (Kg) | Mass ref | NEATM Density (g/cm^3) | +/- (g/cm^3) |
|---|---|---|---|---|---|---|---|---|---|---|---|
| 45 | Eugenia | 210.0 | 31.1 | 0.042 | 0.015 | 1.066 | 0.077 | 5.66E+18 | 1 | 1.17 | 0.17 |
| 130 | Elektra | 174.9 | 25.5 | 0.082 | 0.029 | 0.922 | 0.077 | 6.57E+18 | 3 | 2.34 | 0.34 |
| 762 | Pulcova | 149.4 | 18.7 | 0.039 | 0.013 | 1.076 | 0.108 | 1.40E+18 | 1 | 0.80 | 0.10 |
| 22 | Kalliope | 170.3 | 23.4 | 0.160 | 0.063 | 0.833 | 0.150 | 7.75E+18 | 4 | 3.00 | 0.41 |
| 87 | Sylvia | 272.4 | 40.2 | 0.040 | 0.017 | 1.027 | 0.127 | 1.48E+19 | 5 | 1.40 | 0.21 |
| 107 | Camilla | 241.6 | 35.0 | 0.045 | 0.019 | 1.047 | 0.150 | 1.12E+19 | 1 | 1.52 | 0.22 |
| 121 | Hermione | 165.8 | 22.1 | 0.077 | 0.025 | 0.988 | 0.079 | 5.03E+18 | 2 | 2.11 | 0.28 |
| 216 | Kleopatra | 152.5 | 21.3 | 0.091 | 0.041 | 1.252 | 0.125 | 4.64E+18 | 6 | 2.50 | 0.35 |
| 283 | Emma | 143.9 | 19.2 | 0.028 | 0.011 | 0.829 | 0.053 | 1.38E+18 | 3 | 0.88 | 0.12 |
| 9069 | Hovland | 2.9 | 0.4 | 0.373 | 0.089 | 1.871 | 0.087 | unk | | unk | unk |
| 137170 | 1999 HF1 | 4.4 | 0.6 | 0.155 | 0.059 | 1.784 | 0.048 | unk | | unk | unk |
| 809 | Lundia | 9.6 | 1.1 | 0.144 | 0.031 | 0.850 | 0.041 | 8.10E+14 | this work | 1.77 | 0.20 |
| 854 | Frostia | 9.7 | 1.2 | 0.272 | 0.070 | 0.805 | 0.029 | 3.60E+14 | 7 | 0.90 | 0.20 |
| 3782 | Celle | 6.6 | 0.7 | 0.232 | 0.090 | 1.015 | 0.033 | 3.40E+14 | 8 | 2.40 | 1.10 |
| 1089 | Tama | 12.2 | 1.6 | 0.267 | 0.083 | 0.860 | 0.025 | 2.00E+15 | 7 | 2.50 | 0.40 |
| 1313 | Berna | 13.3 | 1.4 | 0.212 | 0.076 | 0.882 | 0.022 | 1.20E+15 | based on 7 | 1.30 | 0.40 |
| 1333 | Cevenola | 11.2 | 1.4 | 0.214 | 0.081 | 1.162 | 0.038 | unk | | unk | unk |
| 1509 | Esclangona | 9.0 | 1.0 | 0.107 | 0.021 | 1.115 | 0.018 | unk | | unk | unk |
| 3623 | Chaplin | 11.1 | 1.5 | 0.266 | 0.097 | 0.946 | 0.043 | unk | | unk | unk |
| 3749 | Balam | 4.7 | 0.5 | 0.277 | 0.096 | 0.926 | 0.053 | 0.7-2.0E+14 | 9 | 1.3-3.7 | |
| 4674 | Pauling | 4.7 | 0.5 | 0.387 | 0.090 | 0.975 | 0.038 | unk | | unk | unk |
| 5407 | 1992 AX | 3.8 | 0.4 | 0.199 | 0.078 | 1.146 | 0.037 | unk | | unk | unk |
| 5905 | Johnson | 4.1 | 0.5 | 0.266 | 0.100 | 1.133 | 0.056 | unk | | unk | unk |
| 69230 | Hermes | 0.8 | 0.1 | 0.265 | 0.099 | 1.233 | 0.051 | unk | | unk | unk |
| | 1994 AW1 | 1.0 | 0.1 | 0.277 | 0.102 | 1.458 | 0.060 | unk | | unk | unk |
| 4492 | Debussy | 16.5 | 1.9 | 0.039 | 0.018 | 0.907 | 0.011 | 1.50E+15 | 7 | 0.90 | 0.10 |
| 76818 | 2000 RG79 | 2.6 | 0.3 | 0.502 | 0.153 | 1.021 | 0.117 | unk | | unk | unk |
| 185851 | 2000 DP107 | 1.0 | 0.1 | 0.111 | 0.036 | 1.628 | 0.032 | 4.60E+11 | 10 | 0.90 | 0.30 |

**Table 6:** Roughness models used in the TPM analysis; parameters are taken from Muller et al. (1999). $\rho_{crat}$ denotes the area fraction covered in craters, $\gamma_{crat}$ denotes their opening angle: $\gamma_{crat} = 180$ deg for hemispherical craters, lower values for more subdued crater cross-sections.

| Model | $\rho_{crat}$ | $\gamma_{crat}$ |
|---|---|---|
| No roughness | 0 | 0 |
| Low roughness | 40% | 117.7° |
| Nominal roughness | 60% | 144.6° |
| High roughness | 100% | 151.8° |

**Table 7:** Spin-state parameters of the 7 TPM targets as taken from the DAMIT database (see text): rotation period P and J2000 ecliptic longitude λ and latitude β. The two lines for (283) Emma correspond to the two possible spin solutions; both are used in the TPM analysis. SSL and SOL denote the latitude of the sub-solar and sub-observer point, respectively, on the asteroid surface for the time of our Spitzer observations. In the last column, the local time at the sub-observer point is given (local noon at 12 h).

| ID | Asteroid | P | λ | β | SSL | SOL | Sub-observer time |
|----|----------|-----|-----|-----|-------|-------|-------------------|
|    |          | *h* | ° | ° | ° | ° | *h* |
| 107 | Camilla | 4.843928 | 73 | 54 | 15.9 | 23.6 | 13.1 |
| 283 | Emma shape 1 | 6.895221 | 85 | 37 | 55.7 | 47.1 | 13.9 |
| 283 | Emma shape 2 | 6.895222 | 251 | 22 | -55.1 | -39.8 | 13.3 |
| 45 | Eugenia | 5.699143 | 123 | -33 | -34.2 | -45.5 | 10.4 |
| 121 | Hermione | 5.550877 | 359 | 8 | -14 | 3 | 11.9 |
| 130 | Elektra | 5.224664 | 64 | -88 | 3.6 | 4.2 | 13 |
| 22 | Kalliope | 4.1482 | 196 | 3 | 66.7 | 46.6 | 12.4 |
| 87 | Sylvia | 5.18364 | 71 | 66 | 5.1 | 14.3 | 11 |

**Table 8:** TPM results for the seven asteroids with available shape models: equivalent diameter $D_{eq}$, geometric albedo $p_V$, and thermal inertia $\Gamma$. As detailed in the text, NEATM diameters must be lightcurve corrected to facilitate a comparison with the TPM results. Lightcurve correction factors are given in column 6, corrected NEATM diameters (original result from Table 5 times LC Corr) are given in column 7. The comparison in column 8 shows that, after the lightcurve correction, NEATM and TPM diameters agree within the error bars. The *LC corr* factor was used to generated a synthetic TPM lightcurves at 10 μm to calculate the corrected NEATM $D_{eq}$ taking into account the elongated shape of the primary. From the total mass derived by the mutual orbit (column 9), we infer the average density of the system (column 10). The error on the mass (< 7%) is negligible since it is much less than the 30% error on the density introduced by the 10% error on the equivalent diameter.

| ID | Asteroid | $D_{eq}$ (km) | $p_V$ | $\Gamma$ (J s$^{-1/2}$K$^{-1}$m$^{-2}$) | LC corr | corr NEATM D (km) | NEATM/TPM | Mass (kg) | density g/cm$^3$ |
|----|----------|------|------|-----|------|------|------|------|------|
| 22 | Kalliope | 167 ± 17 | 0.166 ± 0.034 | 5—250 | 1.05 | 178.8 | 1.07 | 7.75E+18 | 3.2 ± 0.9 |
| 45 | Eugenia | 198 ± 20 | 0.046 ± 0.010 | 5-85 | 1.01 | 212.1 | 1.07 | 5.66E+18 | 1.4 ± 0.4 |
| 87 | Sylvia | 300 ± 30 | 0.033 ± 0.007 | 10—130 | 1.08 | 294.2 | 0.98 | 1.478E+19 | 1.0 ± 0.3 |
| 107 | Camilla | 245 ± 25 | 0.043 ± 0.009 | 15-35 | 1.06 | 256.0 | 1.05 | 1.12E+19 | 1.5 ± 0.3 |
| 121 | Hermione | 220 ± 22 | 0.043 ± 0.009 | 5-60 | 1.16 | 192.4 | 0.87 | 5.03E+18 | 0.9 ± 0.3 |
| 130 | Elektra | 197 ± 20 | 0.064 ± 0.013 | 5—65 | 1.15 | 201.2 | 1.02 | 6.57E+18 | 1.6 ± 0.5 |
| 283 | Emma shape 1 | 133 ± 14 | 0.032 ± 0.007 | 5-215 | 0.98 | 142.2 | 1.07 | 1.38E+18 | 1.1 ± 0.4 |
| 283 | Emma shape 2 | 137 ± 14 | 0.031 ± 0.007 | 5-130 | 0.96 | 139.3 | 1.02 | 1.38E+18 | 1.0 ± 0.3 |

**Electronic Supplement Tables**

**Table S1:** Intensity, 1-sigma error and contrast of the error in percent of the emissivity spectra for the 28 asteroids of our sample at 7, 11, 20, 30 and 35 μm derived from the spectra shown in Fig. 1a-h. The large errors which were measured for (76818) 2000 RG79, (1333) Cevenola, and (9069) Hovland (contrast >10% at 7 μm) suggest a poor SNR on these spectra, we discarded them from our analysis in Section 5.2.

| Asteroid | Emissivity Intensity | | | | | 1-sigma error | | | | | error/intensity*100 | | | | |
|---|---|---|---|---|---|---|---|---|---|---|---|---|---|---|---|
| | 7 um | 11 um | 20 um | 30 um | 35 um | 7 um | 11 um | 20 um | 30 um | 35 um | 7 um | 11 um | 20 um | 30 um | 35 um |
| 1992_AX | 0.835 | 0.901 | 0.904 | 0.890 | 0.910 | 0.016 | 0.007 | 0.008 | 0.008 | 0.020 | 1.87 | 0.77 | 0.93 | 0.93 | 2.23 |
| 1994_AW1 | 0.904 | 0.902 | 0.914 | 0.907 | 0.955 | 0.037 | 0.011 | 0.011 | 0.013 | 0.061 | 4.10 | 1.21 | 1.19 | 1.41 | 6.42 |
| 1999_HF1 | 0.900 | 0.886 | 0.907 | 0.928 | 0.930 | 0.010 | 0.005 | 0.007 | 0.004 | 0.012 | 1.14 | 0.58 | 0.75 | 0.39 | 1.28 |
| 2000_DP107 | 0.908 | 0.879 | 0.896 | 0.941 | 0.953 | 0.007 | 0.002 | 0.004 | 0.003 | 0.015 | 0.73 | 0.25 | 0.49 | 0.29 | 1.55 |
| 2000_RG79 | 1.068 | 0.899 | 0.853 | 0.810 | 1.107 | 0.249 | 0.029 | 0.040 | 0.036 | 0.136 | 23.34 | 3.27 | 4.71 | 4.49 | 12.33 |
| Balam | 0.778 | 0.910 | 0.923 | 0.893 | 0.967 | 0.048 | 0.006 | 0.007 | 0.007 | 0.037 | 6.12 | 0.61 | 0.78 | 0.82 | 3.86 |
| Berna | 0.807 | 0.929 | 0.889 | 0.885 | 0.907 | 0.005 | 0.003 | 0.004 | 0.003 | 0.012 | 0.66 | 0.30 | 0.43 | 0.36 | 1.33 |
| Camilla | 0.828 | 0.930 | 0.903 | 0.950 | 0.928 | 0.007 | 0.006 | 0.002 | 0.003 | 0.006 | 0.86 | 0.61 | 0.24 | 0.26 | 0.66 |
| Celle | 0.703 | 0.919 | 0.885 | 0.913 | 1.018 | 0.037 | 0.006 | 0.006 | 0.006 | 0.018 | 5.21 | 0.62 | 0.66 | 0.67 | 1.75 |
| Cevenola | 0.753 | 0.943 | 0.892 | 0.946 | 0.954 | 0.109 | 0.010 | 0.008 | 0.009 | 0.036 | 14.51 | 1.11 | 0.93 | 0.91 | 3.77 |
| Chaplin | 0.722 | 0.918 | 0.927 | 0.892 | 0.942 | 0.050 | 0.010 | 0.007 | 0.008 | 0.021 | 6.96 | 1.12 | 0.77 | 0.92 | 2.20 |
| Debussy | 0.808 | 0.911 | 0.925 | 0.882 | 0.899 | 0.006 | 0.001 | 0.003 | 0.003 | 0.007 | 0.73 | 0.16 | 0.31 | 0.29 | 0.82 |
| Elektra | 0.732 | 0.902 | 0.903 | 0.861 | 0.866 | 0.009 | 0.005 | 0.005 | 0.008 | 0.011 | 1.17 | 0.57 | 0.54 | 0.92 | 1.21 |
| Emma | 0.844 | 0.953 | 0.906 | 0.935 | 0.920 | 0.011 | 0.010 | 0.002 | 0.002 | 0.009 | 1.33 | 1.04 | 0.25 | 0.25 | 0.95 |
| Esclangona | 0.800 | 0.922 | 0.909 | 0.906 | 0.826 | 0.008 | 0.003 | 0.004 | 0.004 | 0.026 | 0.94 | 0.36 | 0.48 | 0.44 | 3.20 |
| Eugenia | 0.828 | 0.934 | 0.932 | 0.929 | 0.936 | 0.008 | 0.004 | 0.003 | 0.003 | 0.007 | 1.00 | 0.45 | 0.28 | 0.29 | 0.74 |
| Frostia | 0.844 | 0.898 | 0.905 | 0.901 | 0.892 | 0.017 | 0.008 | 0.008 | 0.011 | 0.032 | 2.03 | 0.90 | 0.84 | 1.18 | 3.54 |
| Hermes | 0.846 | 0.877 | 0.925 | 0.902 | 0.993 | 0.014 | 0.011 | 0.008 | 0.006 | 0.017 | 1.67 | 1.25 | 0.88 | 0.63 | 1.73 |
| Hermione | 0.825 | 0.910 | 0.909 | 0.908 | 0.954 | 0.022 | 0.007 | 0.003 | 0.002 | 0.009 | 2.67 | 0.77 | 0.37 | 0.27 | 0.89 |
| Hovland | 0.739 | 0.886 | 0.912 | 0.891 | 0.828 | 0.137 | 0.018 | 0.013 | 0.016 | 0.045 | 18.50 | 2.01 | 1.48 | 1.76 | 5.45 |
| Johnson | 0.807 | 0.917 | 0.903 | 0.891 | 0.923 | 0.053 | 0.017 | 0.012 | 0.014 | 0.043 | 6.61 | 1.82 | 1.28 | 1.55 | 4.68 |
| Kalliope | 0.718 | 0.874 | 0.921 | 0.918 | 0.938 | 0.003 | 0.003 | 0.002 | 0.002 | 0.006 | 0.43 | 0.38 | 0.25 | 0.26 | 0.69 |
| Lundia | 0.759 | 0.912 | 0.881 | 0.875 | 0.929 | 0.017 | 0.004 | 0.004 | 0.006 | 0.017 | 2.28 | 0.44 | 0.94 | 0.71 | 1.83 |
| Pauling | 0.812 | 0.895 | 0.911 | 0.904 | 0.928 | 0.028 | 0.005 | 0.007 | 0.005 | 0.022 | 3.47 | 0.57 | 0.77 | 0.59 | 2.35 |
| Pulcova | 0.803 | 0.944 | 0.887 | 0.850 | 0.854 | 0.006 | 0.005 | 0.003 | 0.002 | 0.003 | 0.79 | 0.53 | 0.32 | 0.22 | 0.40 |
| Sylvia | 0.862 | 0.926 | 0.910 | 0.894 | 0.897 | 0.009 | 0.017 | 0.003 | 0.004 | 0.005 | 1.03 | 1.84 | 0.36 | 0.48 | 0.53 |
| Tama | 0.797 | 0.908 | 0.911 | 0.894 | 0.903 | 0.008 | 0.003 | 0.005 | 0.005 | 0.015 | 1.03 | 0.30 | 0.55 | 0.51 | 1.70 |

**Table S2a-z,a:** For each 28 asteroids of our survey, we provide an ascii table containing the wavelength of observations (column 1), the flux in *Jy* (column 2), the 3-sigma error on the flux (column 3), the fit by NEATM in *Jy* (colum 4), the emissivity spectrum obtained by dividing the observed flux by the NEATM, and the emissivity error.

An example of the *Kalliope.IRS.4error.2.stm.emi* file is attached below. The files are available electronically.

Object: Kalliope

Dmin = 170.253
eta = 0.833000
pv = 0.160300

| Wavelength[um] | Flux[Jy] | Flux_err[Jy] | NEATM[Jy] | Emissivity | Emis_err |
|-------------|----------|------------|------|-------------|------------|
| 5.21188 | 0.264532 | 0.00579000 | 0.297354 | 0.800658 | 0.0175246 |
| 5.24282 | 0.275568 | 0.00194000 | 0.312910 | 0.792596 | 0.00557988 |
| 5.27374 | 0.292418 | 0.00039000 | 0.329041 | 0.799828 | 0.00106674 |
| 5.30468 | 0.306055 | 0.00184000 | 0.345781 | 0.796601 | 0.00478916 |
| 5.33561 | 0.320093 | 0.00227000 | 0.363125 | 0.793346 | 0.00562616 |
| 5.36654 | 0.324215 | 0.00190000 | 0.381090 | 0.765681 | 0.00448713 |

……

**FIGURES**

**Figure 1a-g:** The left panel shows the spectra or spectral energy distribution (SED) of 28 targets observed with Spitzer/IRS between 5.2 and 38 μm along with the best-fit NEATM thermal model (see Section 3) and the best-fit TPM (see Section 4). The NEATM models are calculated with STM using parameters listed in Table 5. The TPM curves, where available, use the best-fit parameters given in Table 8. The TPM fits are estimated with different degrees of roughness as given in Table 6, geometry of the systems listed in Table 7 and provide the best-fit parameters shown in Table 8. The emissivity spectra created by dividing the measured SED by the best-fit NEATM of each asteroid is displayed on the right panel. The emissivity spectra were rebinned to a resolution of 60 to improve the SNR. The gray boxes superimposed on the emissivity spectra show the location of an instrumental artifact.

**Figure 1a:**

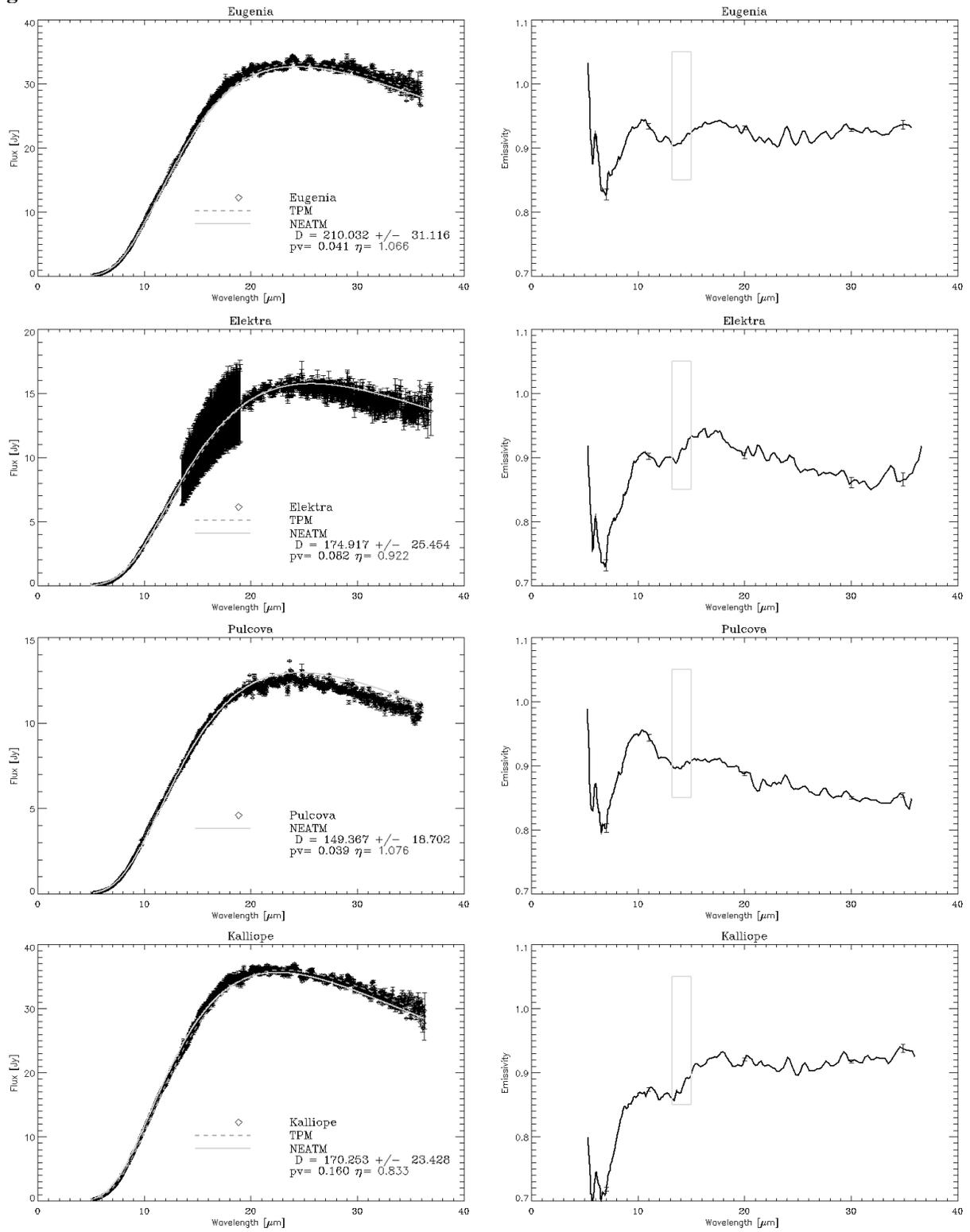

**Figure 1b**:

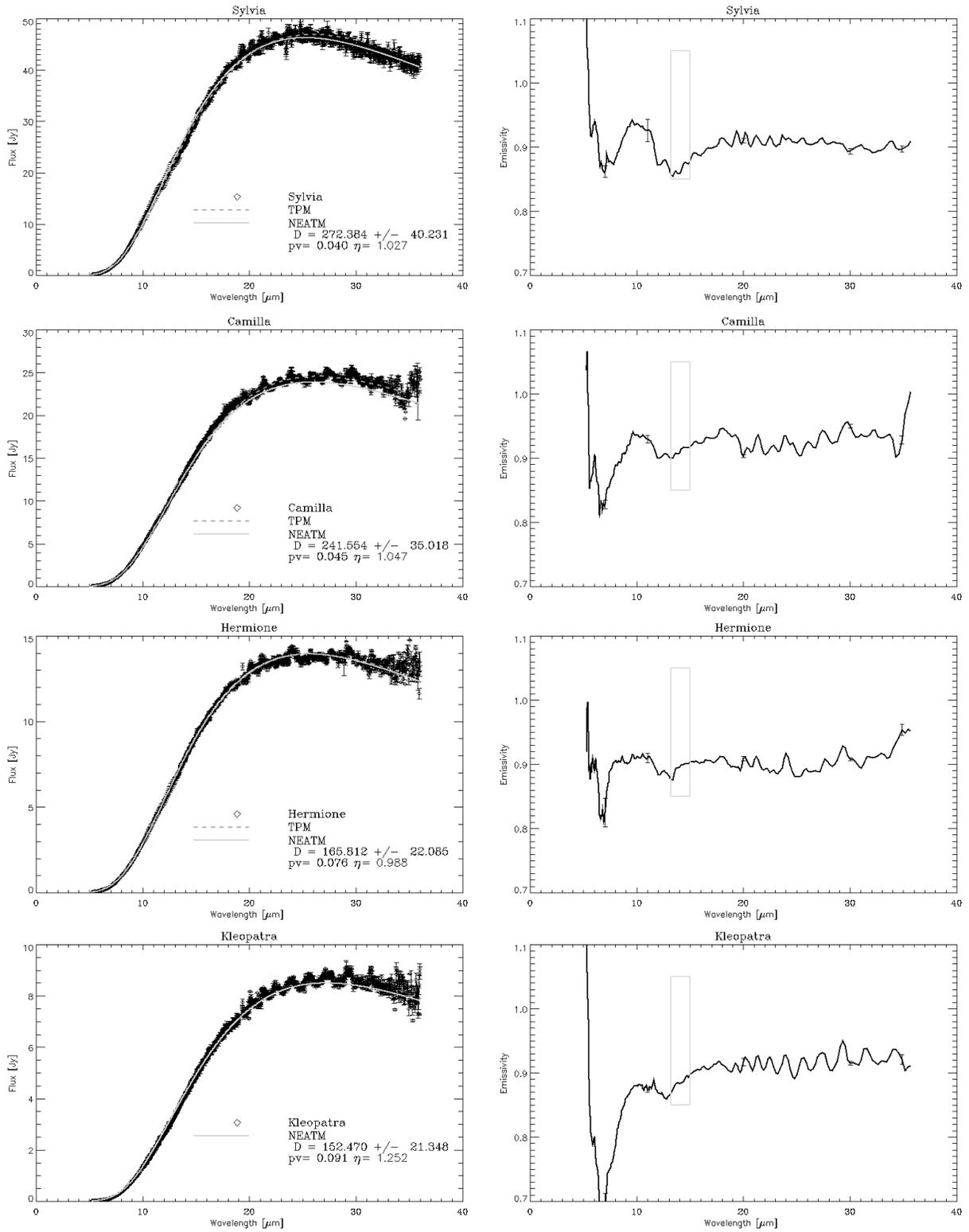

**Figure 1c:**

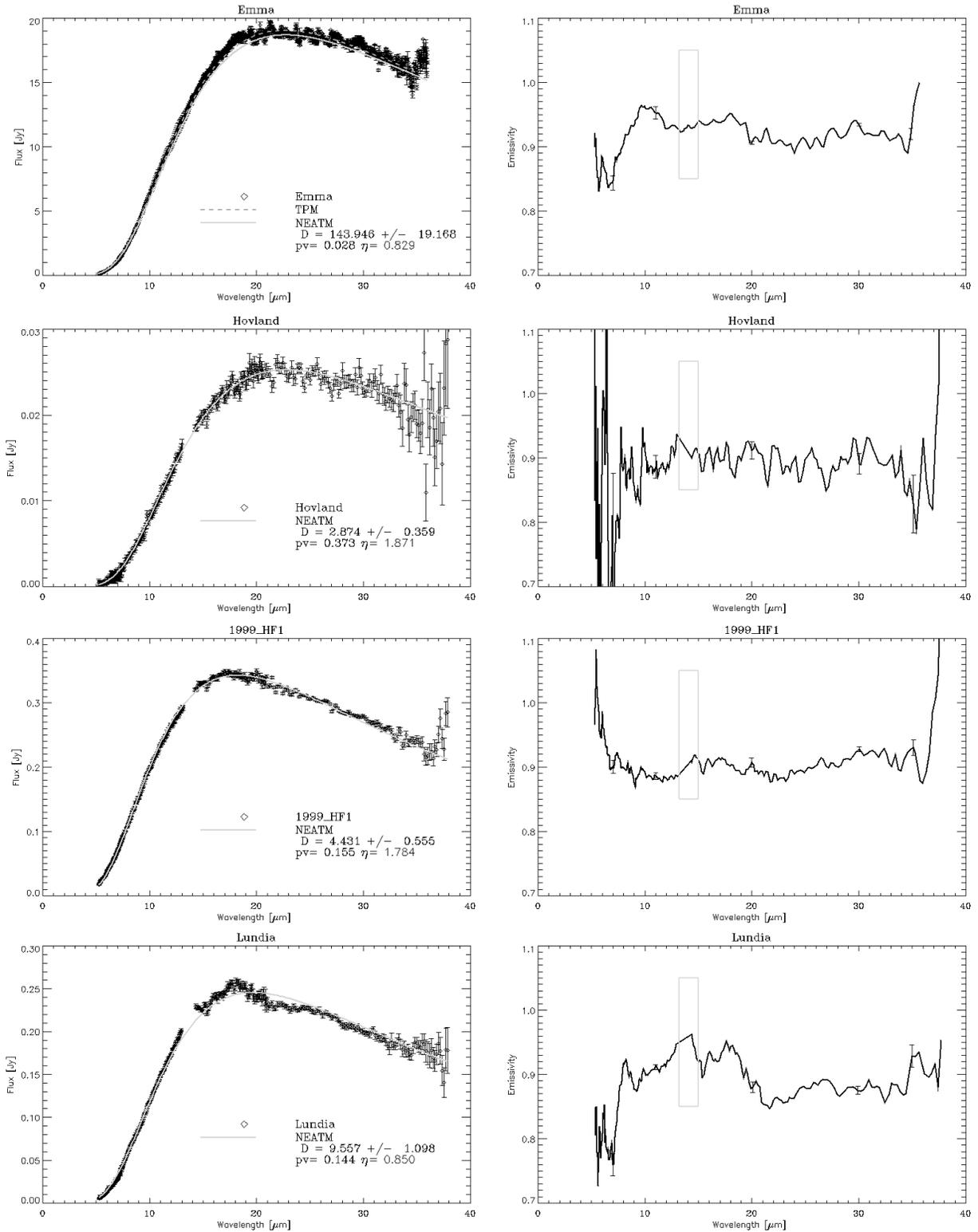

**Figure 1d:**

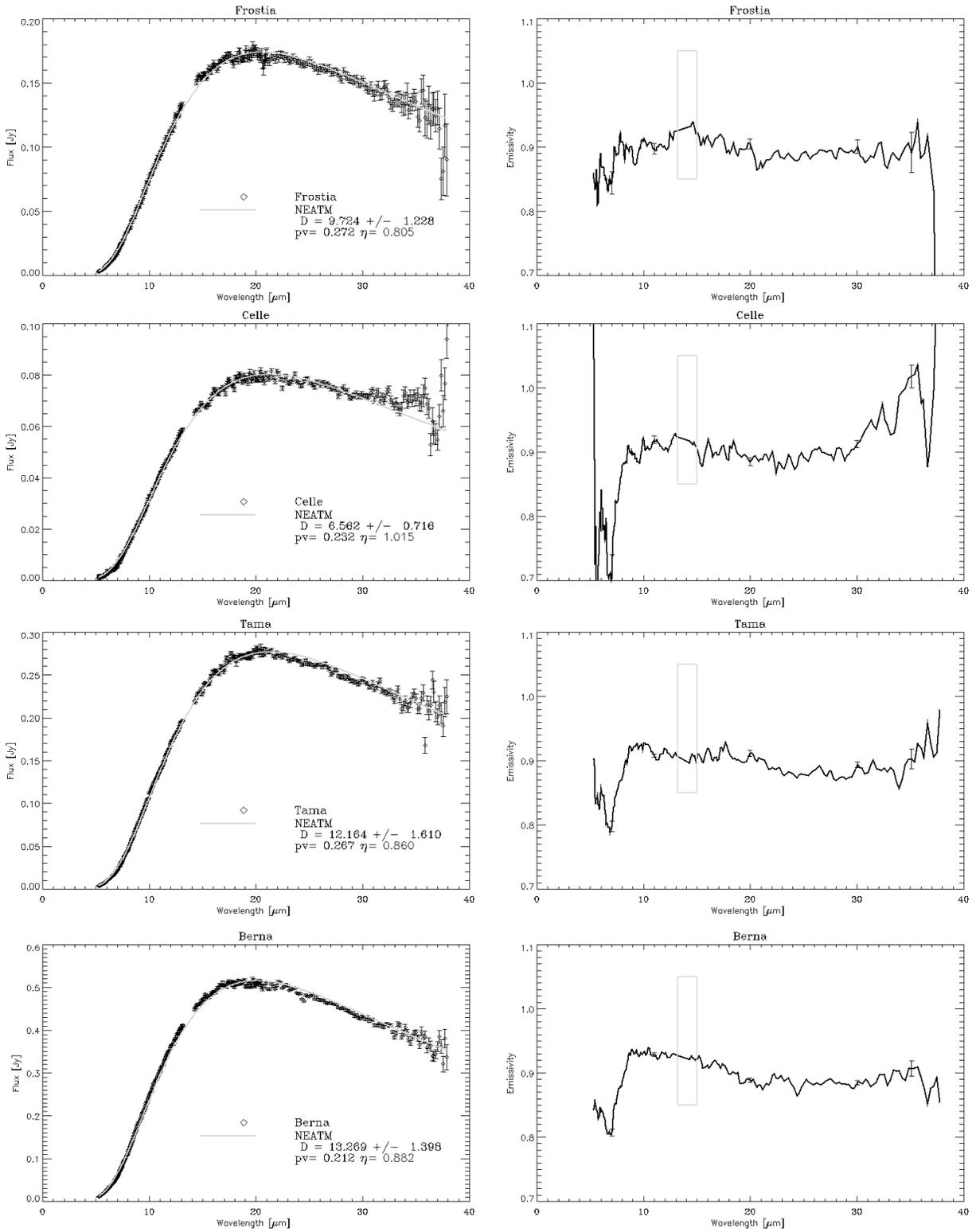

**Fig 1e:**

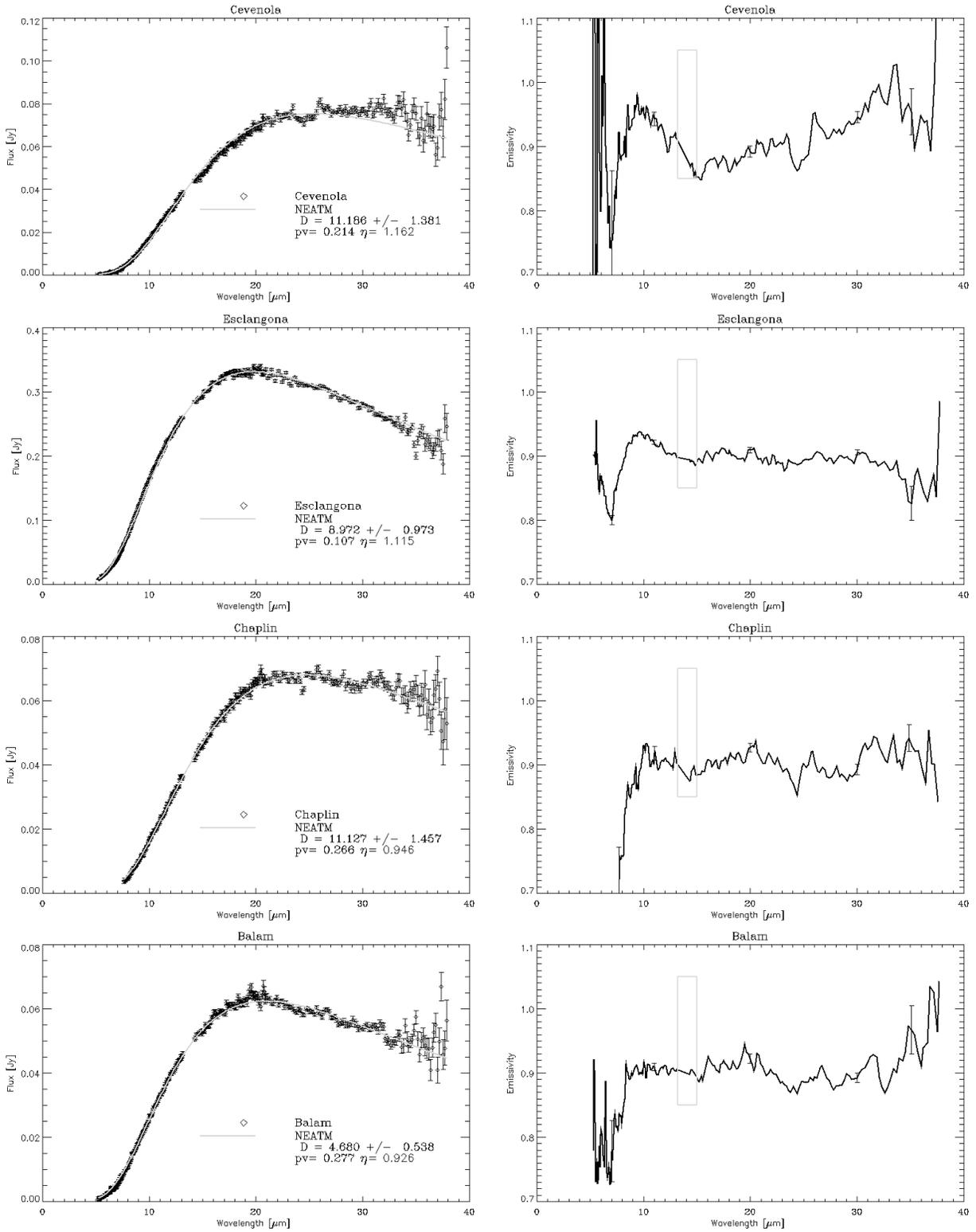

**Fig 1f:**

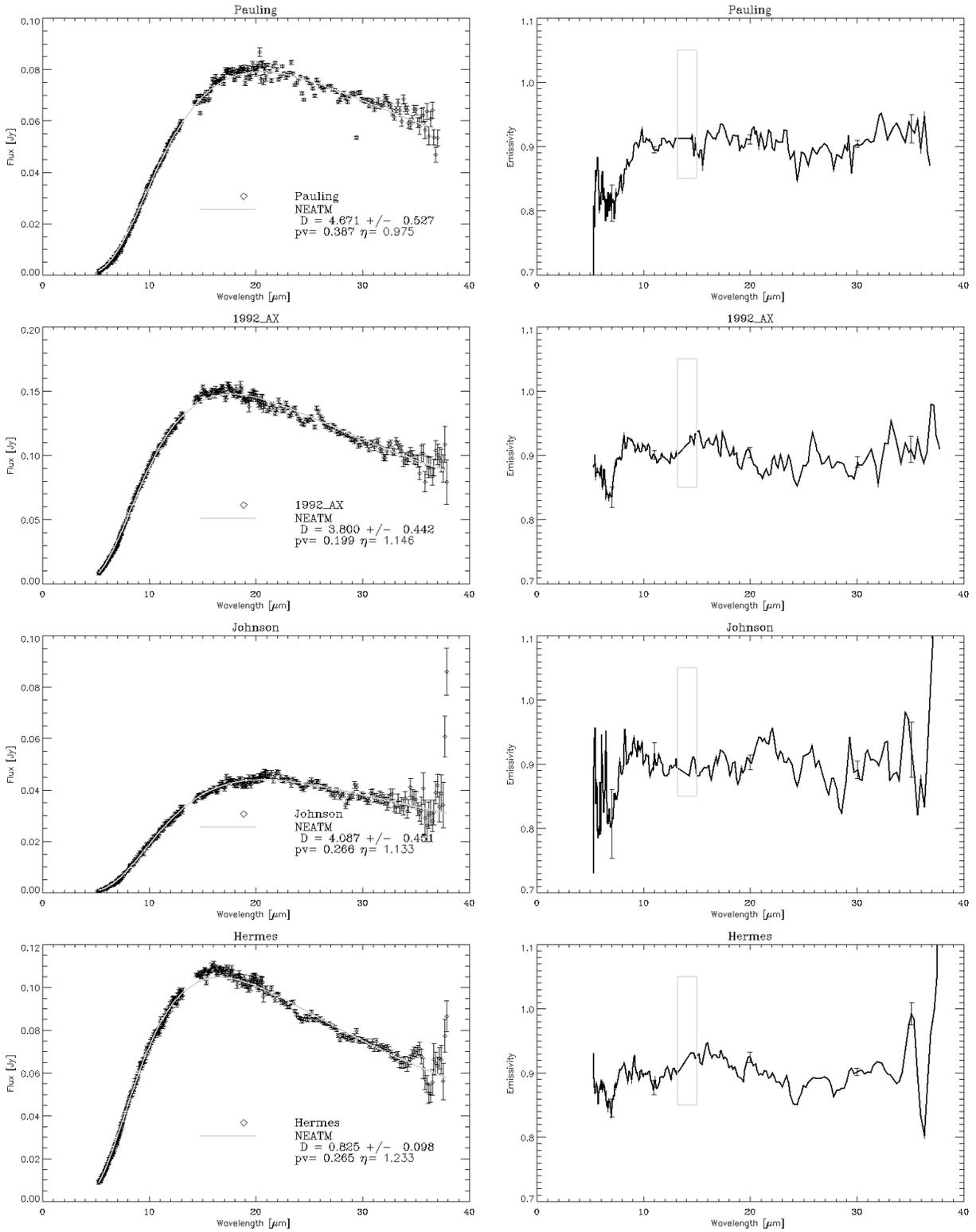

**Fig 1g:**

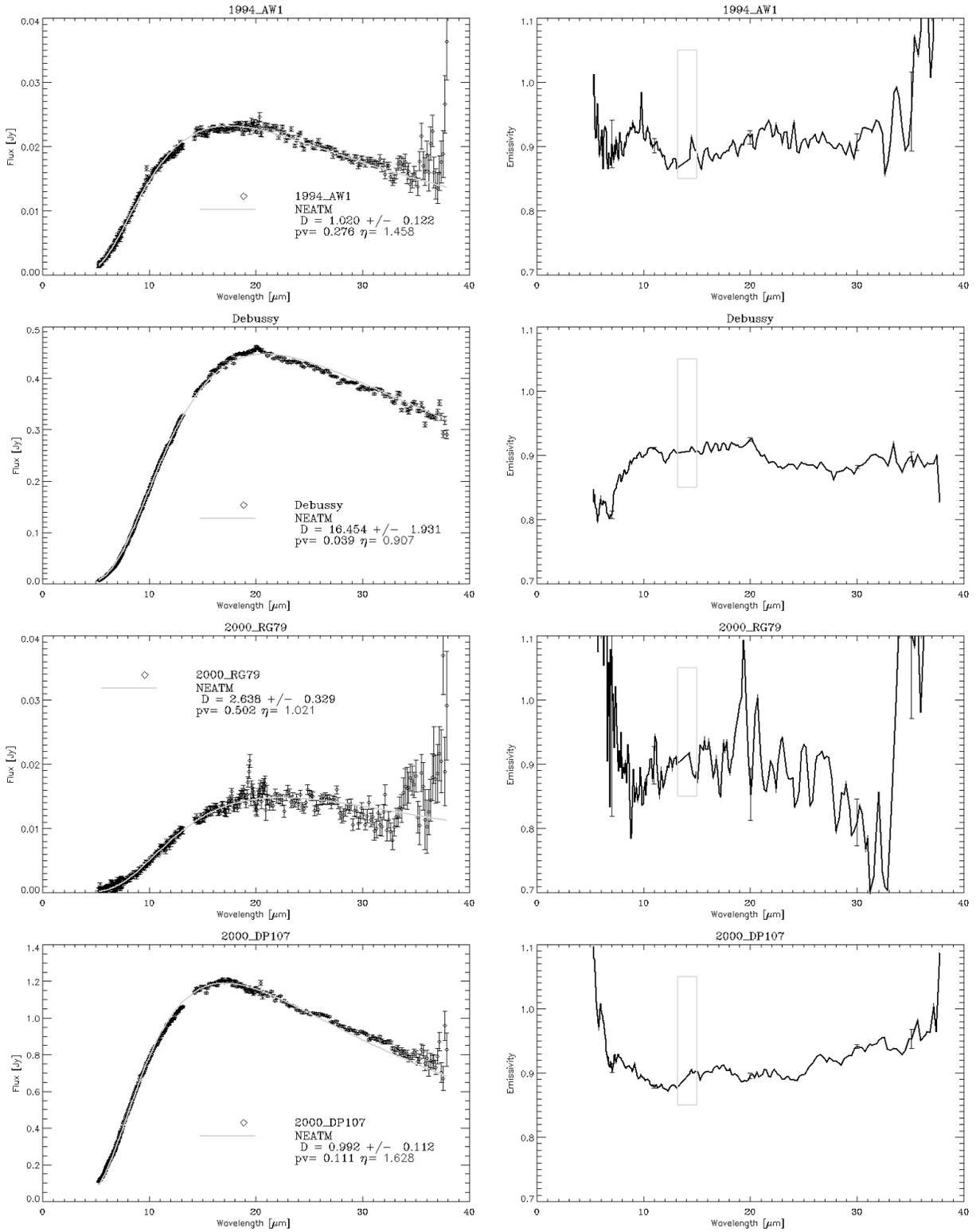

**Figure 2:** Photometric lightcurves of 14 multiple asteroids from our target sample collected 0.5 to 87 days before/after the Spitzer/IRS observations. The photometric variations on these lightcurves which vary from 0.09 to 1.15 mag (average of 0.47 mag) are mostly due to spin of the irregularly shaped larger component of the system. The photometric lightcurves are phased to the rotation period of the primary, the time of the Spitzer/IRS observation is indicated by a dashed line. On the x-axis labeled "fraction of primary rotation", the 0 is defined as the rotation phase during the Spitzer observation. Because the dashed line avoids the deep attenuation signature of a mutual event, the system was not seen during eclipse/occultation/transit. The spin periods of the primary and secondary from VOBAD are labeled; the $\Delta$mag of the lightcurves was also calculated. The text in Section 2.3 describes how we used these lightcurves to derive the absolute magnitude and its error for these asteroids.

**Figure 2a:**

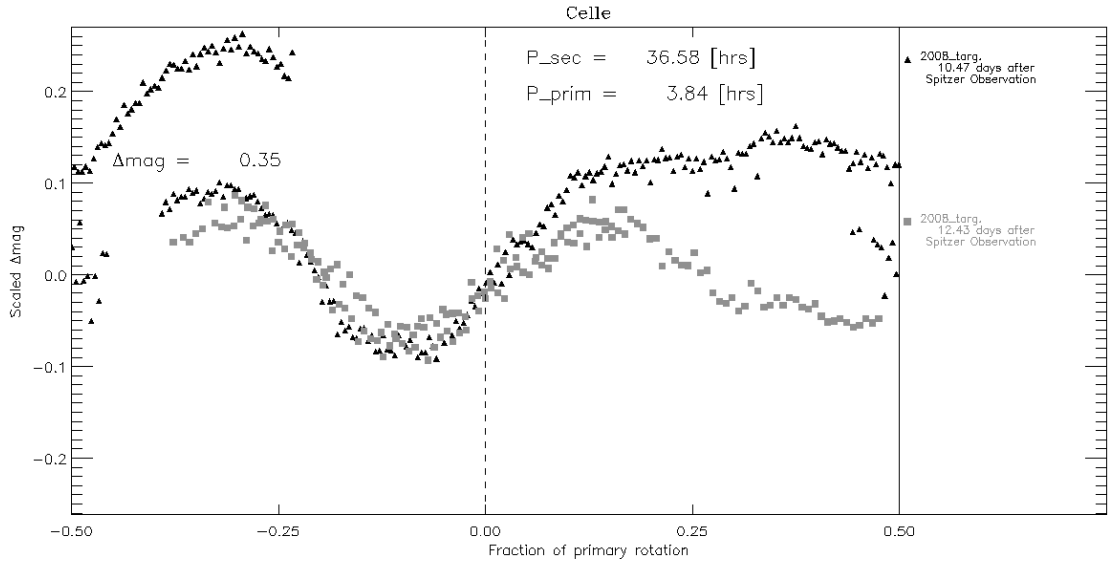

**Celle**

P_sec = 36.58 [hrs]
P_prim = 3.84 [hrs]

2008_targ,
10.47 days after
Spitzer Observation

2008_targ,
12.43 days after
Spitzer Observation

Δmag = 0.35

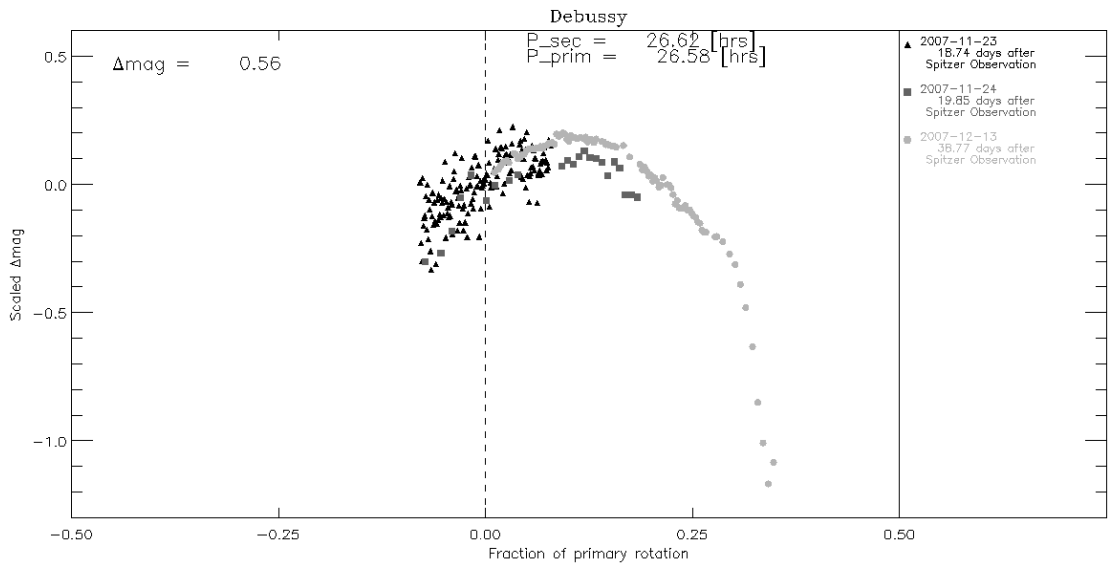

**Debussy**

Δmag = 0.56

P_sec = 26.62 [hrs]
P_prim = 26.58 [hrs]

2007-11-23
18.74 days after
Spitzer Observation

2007-11-24
19.85 days after
Spitzer Observation

2007-12-13
38.77 days after
Spitzer Observation

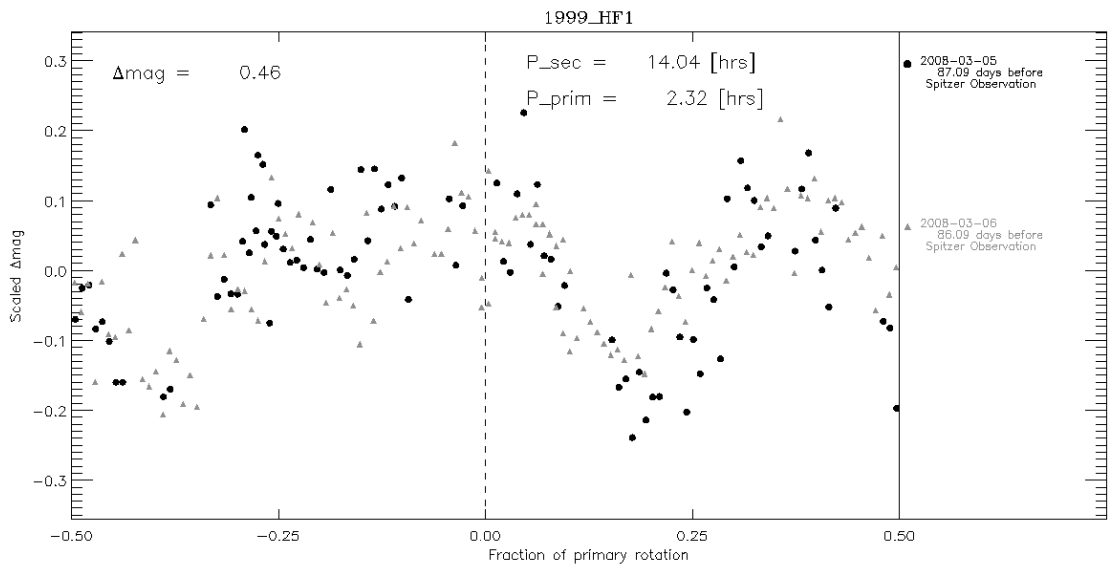

**1999_HF1**

Δmag = 0.46

P_sec = 14.04 [hrs]
P_prim = 2.32 [hrs]

2008-03-05
87.09 days before
Spitzer Observation

2008-03-06
86.09 days before
Spitzer Observation

**Figure 2b:**

### Esclangona

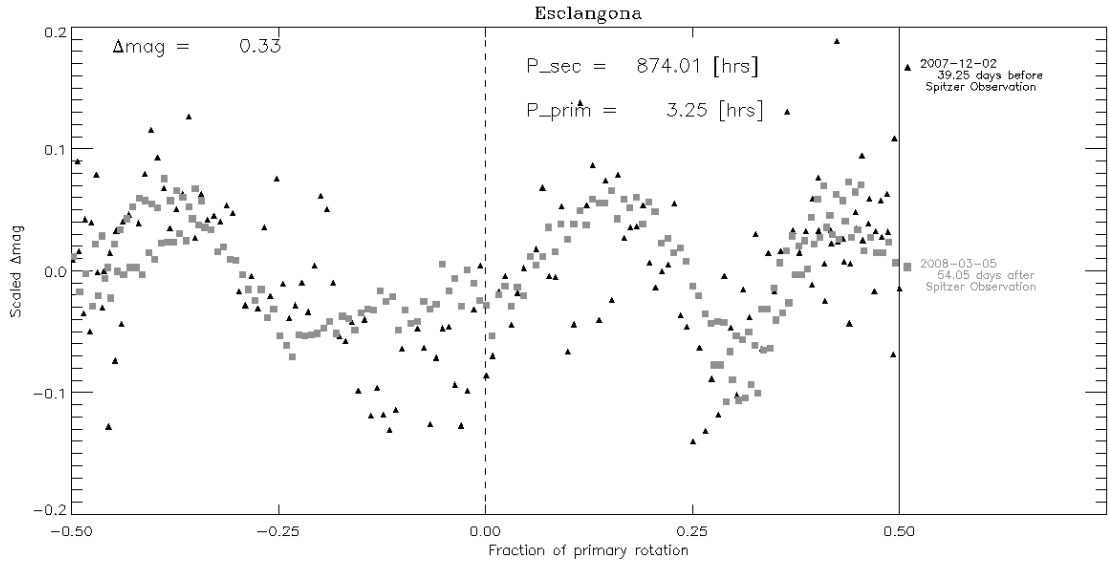

### Berna

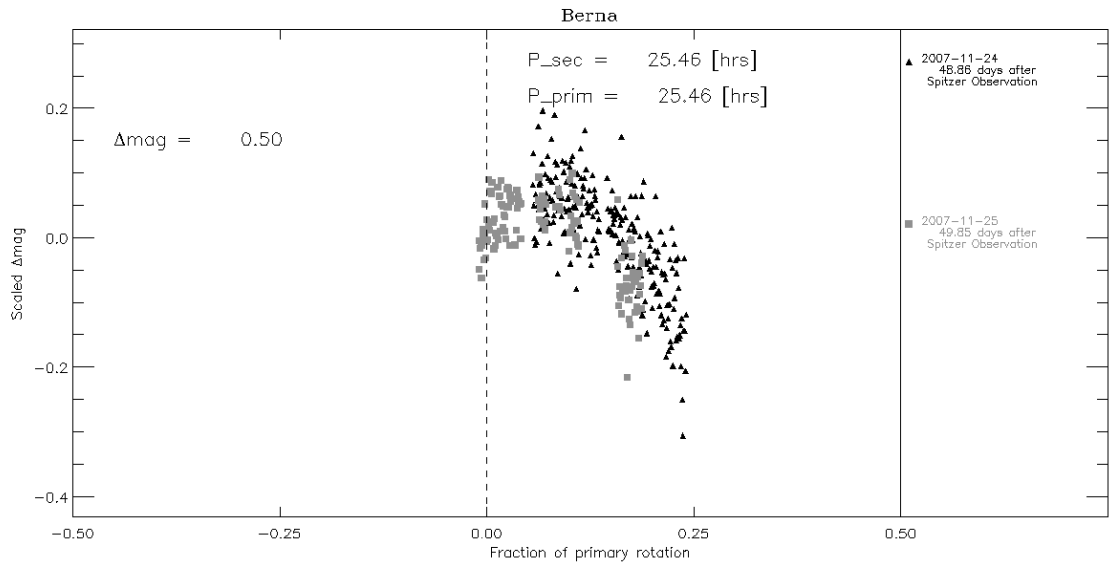

### Cevenola

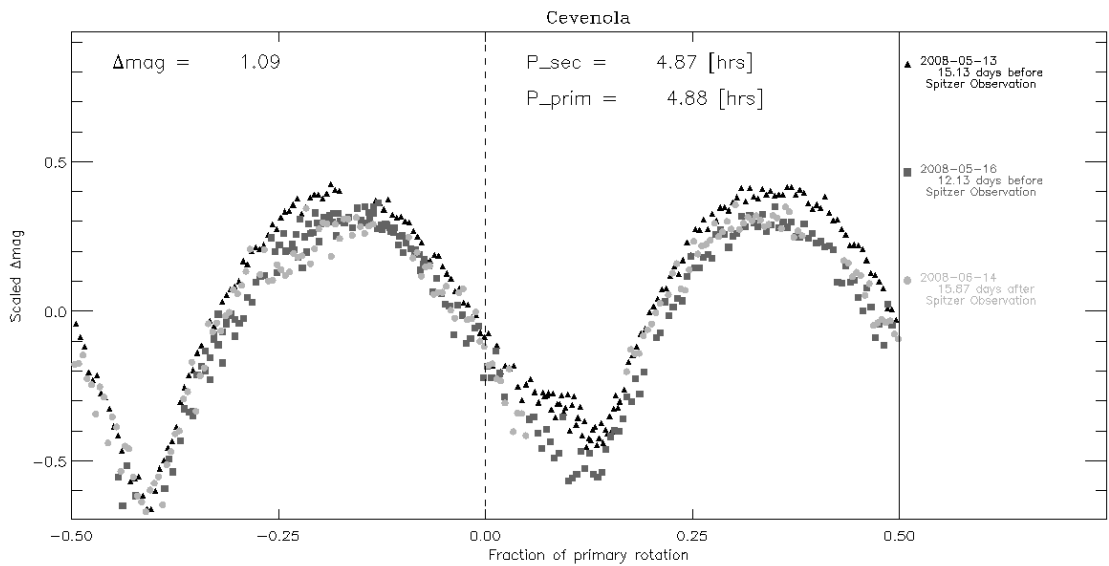

**Figure 2c:**

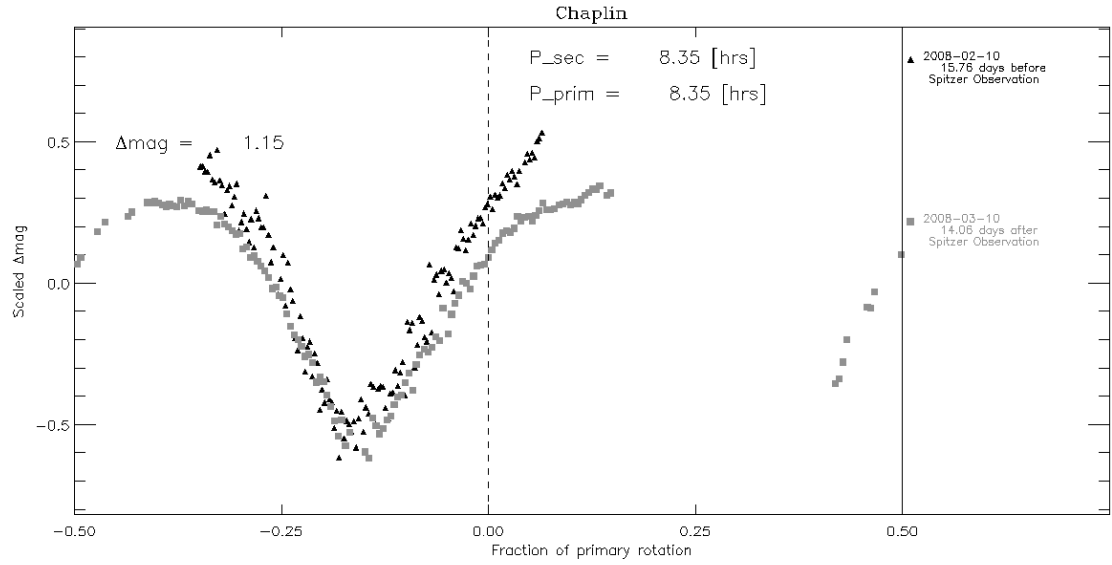

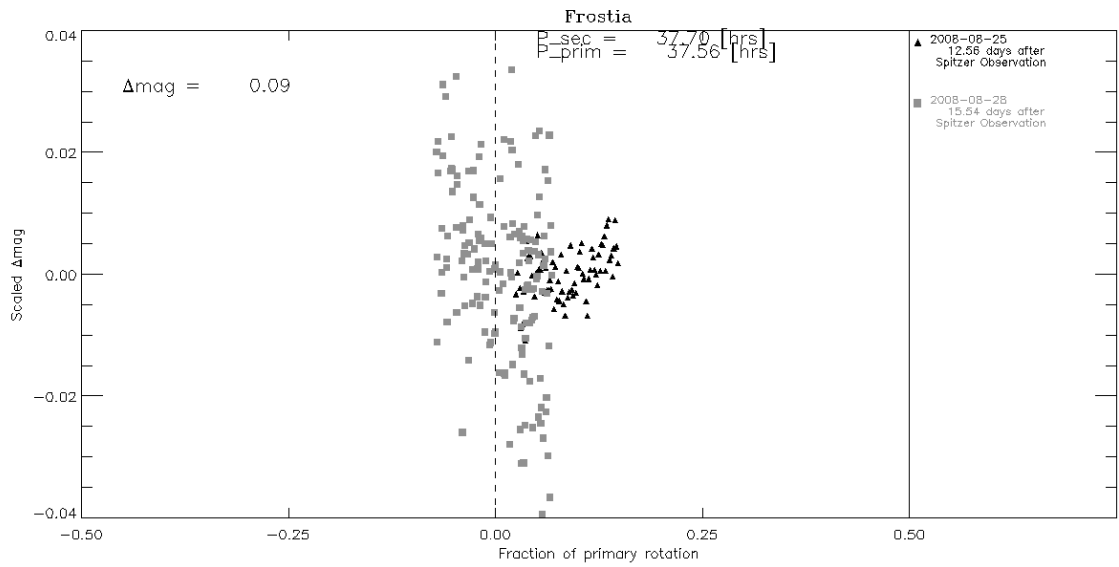

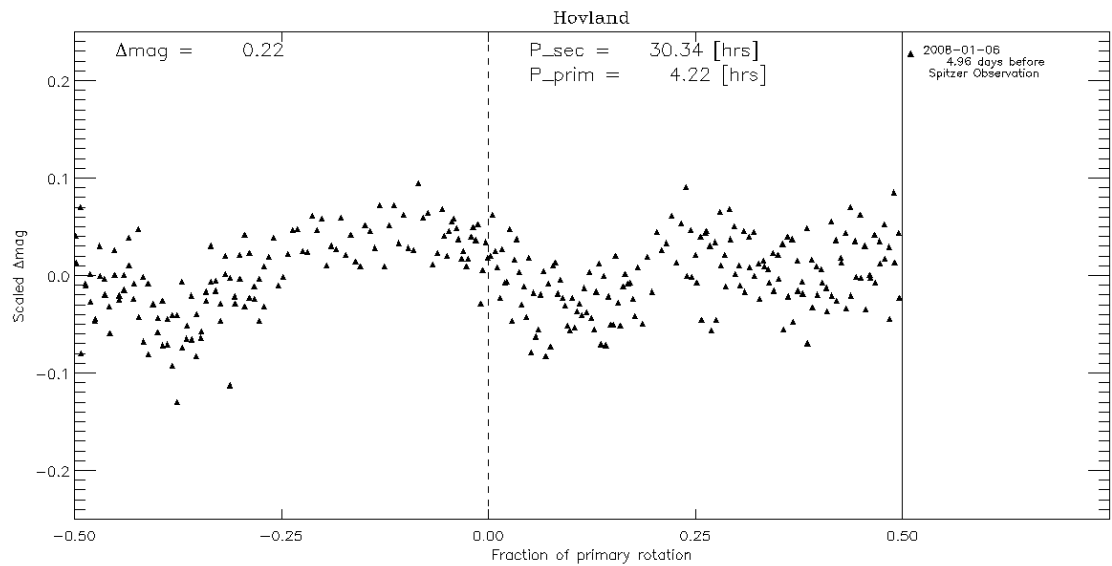

**Figure 2d:**

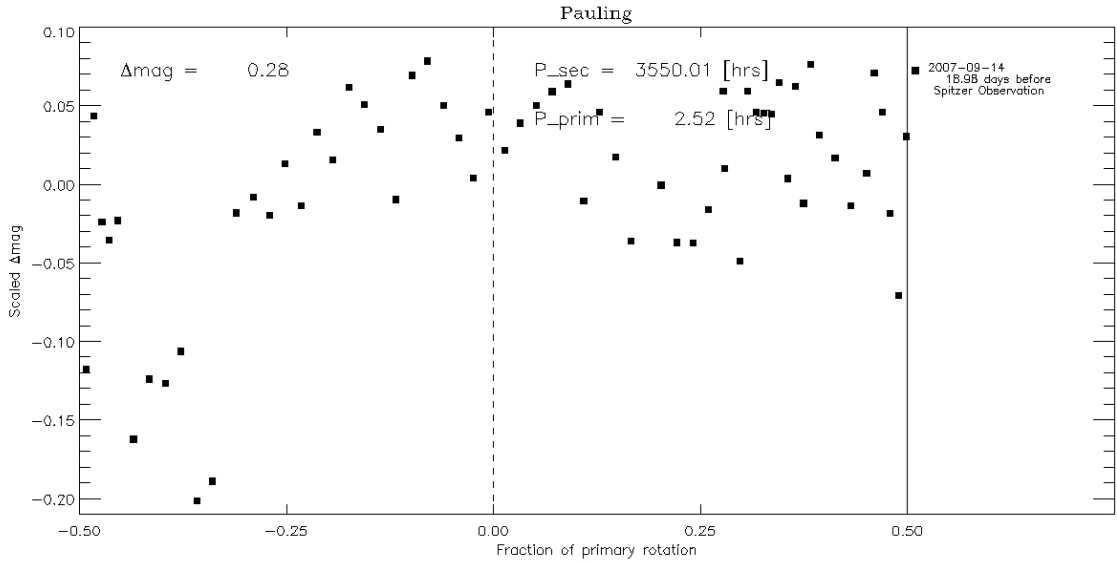

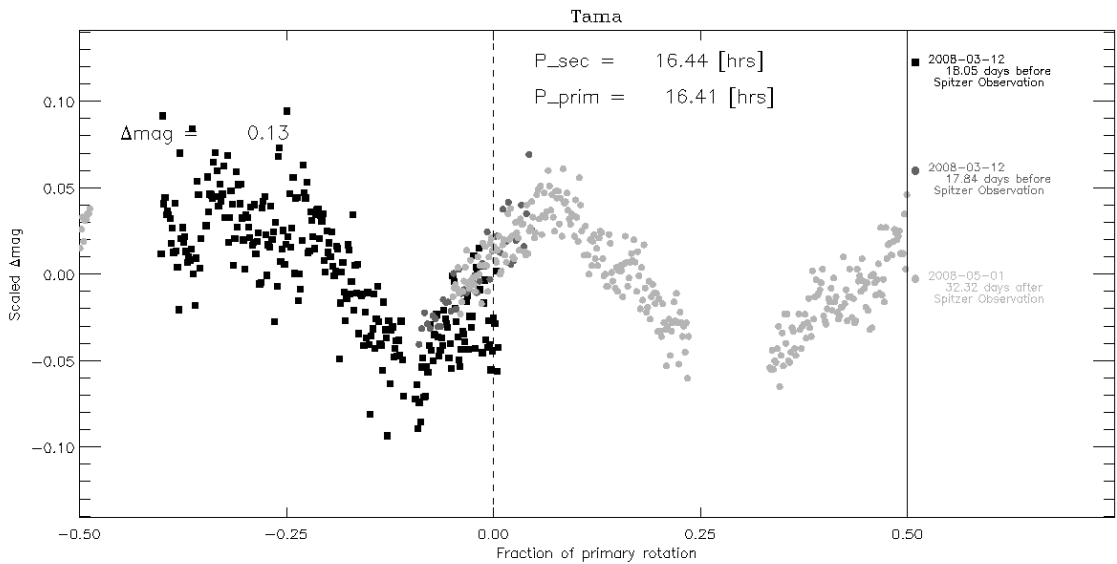

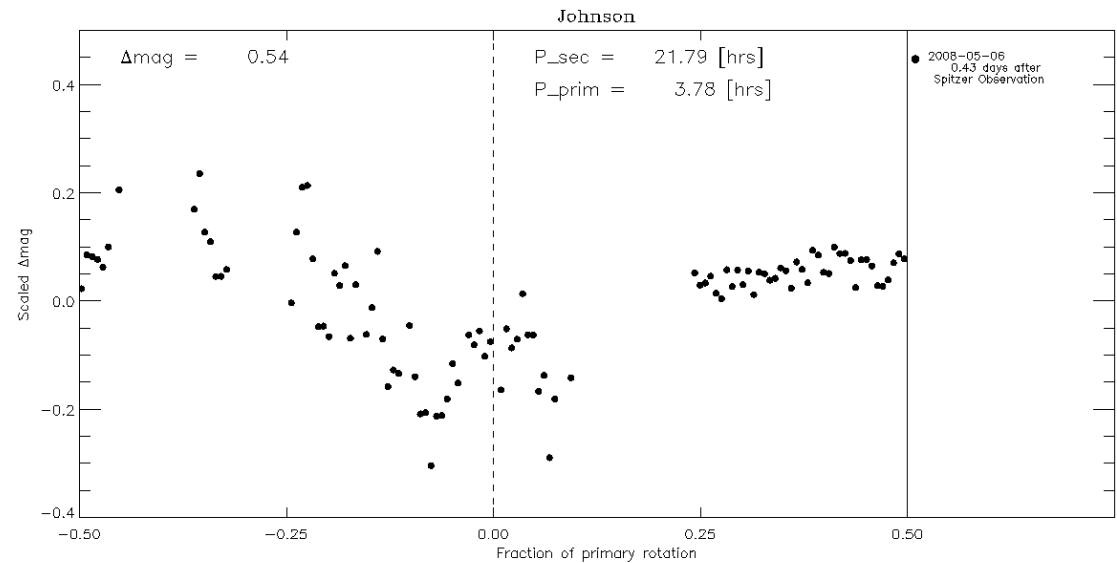

**Figure 2e:**

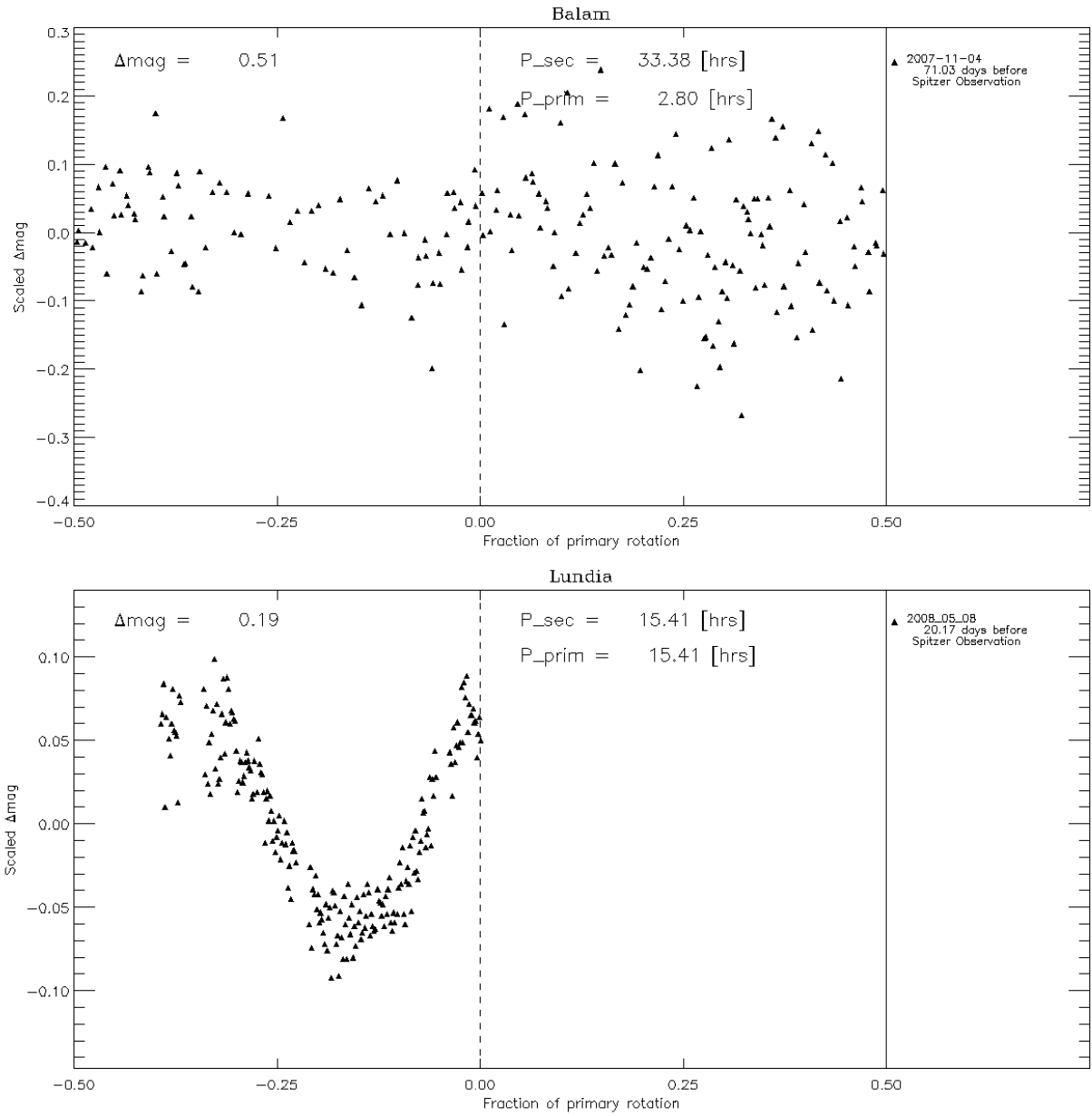

**Figure 3:** TPM fit to IRS observations of (121) Hermione: optimum $\chi^2$ as a function of $\Gamma$ (in J s$^{-1/2}$K$^{-1}$m$^{-2}$) and roughness model (as defined in Table 6). For each $\Gamma$ value, the best-fit $D_{eq}$ and $p_V$ are found (not plotted). The best-fit thermal inertia increases with the assumed surface roughness.

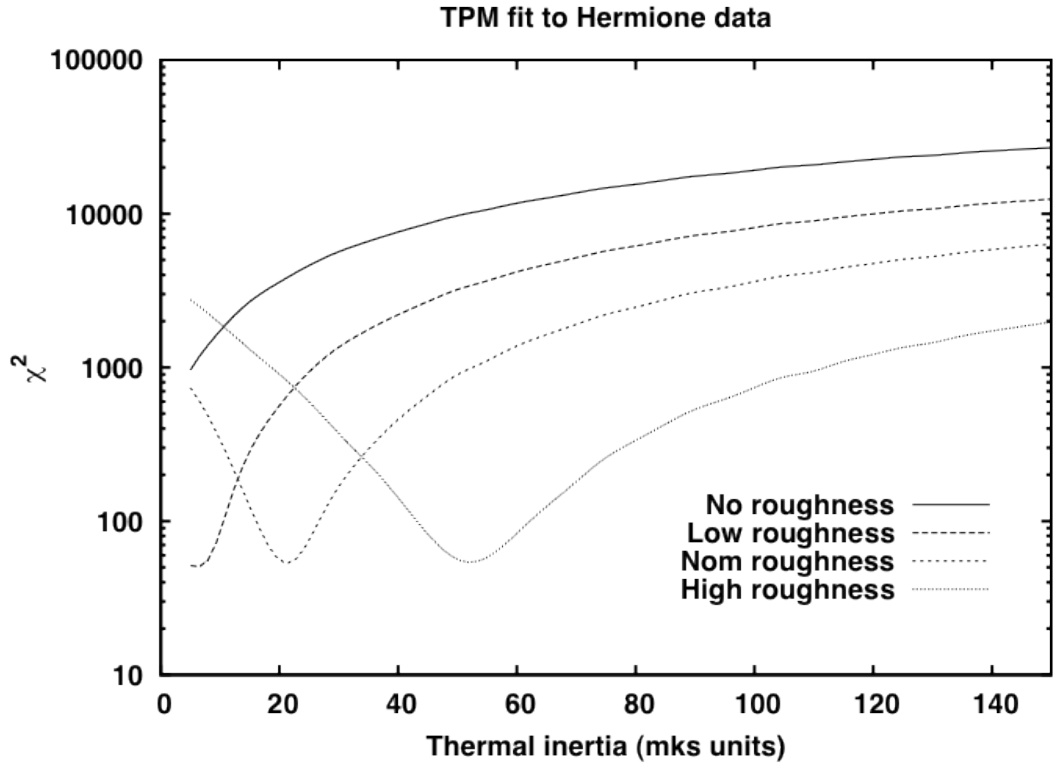

**Figure 4:** Emissivity spectra from 7 to 35 μm of 25 multiple asteroids binned at a resolution of ~50 at 20 μm and sorted by taxonomic class as listed in Table 2. Two asteroids are not yet classified due to the absence of good quality visible/nir NIR reflectance spectra. The dashed lines indicate the wavelength range (between 8.75 and 11.75 μm) where the Christiansen feature and the restrahlen bands are expected. The number in the brackets is the geometric albedo $p_v$ derived from our modeling. We labeled the 1-sigma error at ~13-20 μm generated based on the method described in Section 3.2.

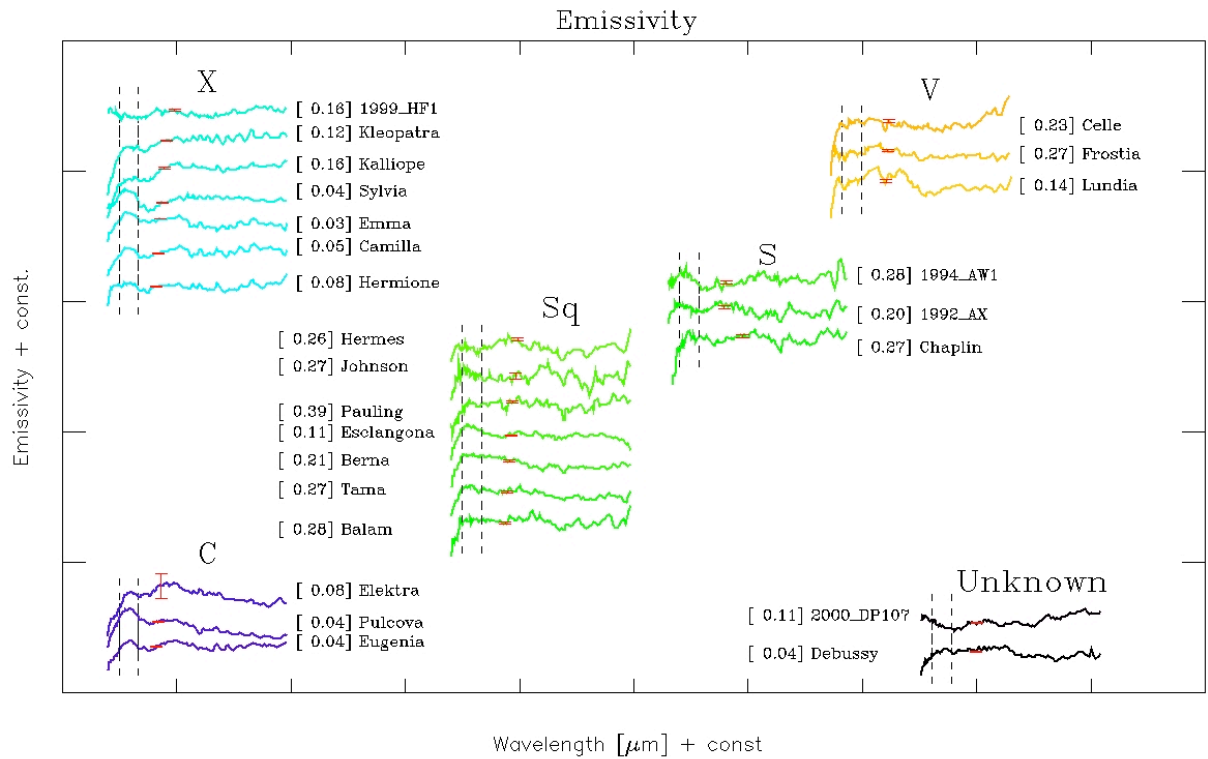

**Figure 5:** Profile of the emissivity spectra (R=50) of our asteroid samples with respect to their radiometric size. The short wavelength emissivity bands from 8.75 and 11.75 μm (dashed line) are detected for all asteroids larger than 17 km but (22) Kalliope & (216) Kleopatra. A 1-sigma error bar is labeled on each spectrum between 15, and 25 μm. The box superimposed on the emissivity spectra shows the location of an instrumental artifact.

# Emissivity by Size

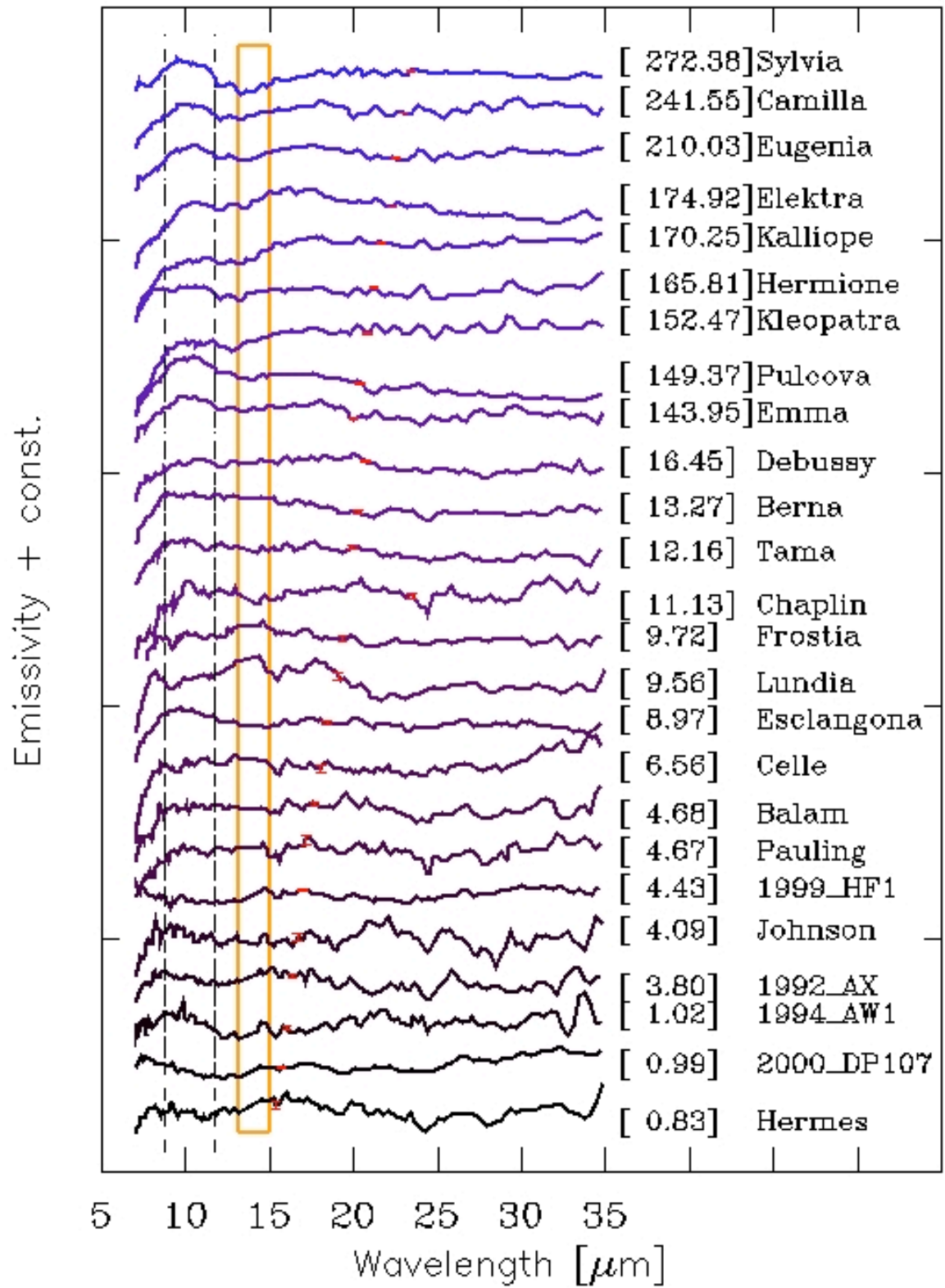

[ 272.38]Sylvia
[ 241.55]Camilla
[ 210.03]Eugenia
[ 174.92]Elektra
[ 170.25]Kalliope
[ 165.81]Hermione
[ 152.47]Kleopatra
[ 149.37]Pulcova
[ 143.95]Emma
[ 16.45] Debussy
[ 13.27] Berna
[ 12.16] Tama
[ 11.13] Chaplin
[ 9.72] Frostia
[ 9.56] Lundia
[ 8.97] Esclangona
[ 6.56] Celle
[ 4.68] Balam
[ 4.67] Pauling
[ 4.43] 1999_HF1
[ 4.09] Johnson
[ 3.80] 1992_AX
[ 1.02] 1994_AW1
[ 0.99] 2000_DP107
[ 0.83] Hermes

Emissivity + const.

Wavelength [$\mu m$]